\documentclass[a4paper,11pt]{article}
\pdfoutput=1
\usepackage{jcappub} % for details on the use of the package, please see the JINST-author-manual
\usepackage{lineno}
\pdfoutput=1 % if your are submitting a pdflatex (i.e. if you have
\usepackage[T1]{fontenc} % if needed
\usepackage{graphicx}
\usepackage{amsmath}
\usepackage{mathtools}
\usepackage{nicefrac}
\usepackage[dvipsnames]{xcolor}
\usepackage{booktabs}
\usepackage[normalem]{ulem}
\usepackage{physics}
\usepackage{enumitem}
\usepackage{bibentry}
\usepackage{pdfpages}
\newcommand{\Msun}{\, h^{-1} \,  M_{\odot}}
\newcommand{\hompc}{\,h\,{\rm Mpc}^{-1}}
\newcommand{\mpcoh}{\,h^{-1}\,{\rm Mpc}}

\NewDocumentCommand{\codeword}{v}{%
\texttt{#1}}%\texttt{\textcolor{blue}{#1}}
\newcommand{\NP}[0]{N_{\rm part}^{1/3}}
\newcommand{\NM}[0]{N_{\rm mesh}^{1/3}}
\newcommand{\NS}[0]{N_{\rm step}}
\newcommand{\LF}[0]{\ell_{\rm Force}}

\newcommand{\nbody}[1]{{\it N}-body}

\title{Fast production of cosmological emulators in modified gravity: the matter power spectrum}

\author[a,b,1]{Bartolomeo Fiorini,\note{Corresponding author.}}
\author[b]{Kazuya Koyama,}
\author[a,b]{Tessa Baker}

\affiliation[a]{Astronomy Unit, Queen Mary University of London, Mile End Road, London E1 4NS, UK}
\affiliation[b]{Institute of Cosmology \& Gravitation, University of Portsmouth, Dennis Sciama Building, Burnaby Road, Portsmouth, PO1 3FX, United Kingdom}
\emailAdd{bartolomeo.fiorini@port.ac.uk}

\abstract{We test the convergence of fast simulations based on the COmoving Lagrangian Acceleration (COLA) method for predictions of the matter power spectrum, specialising our analysis in the redshift range $1 \le z \le 1.65$, relevant to high-redshift spectroscopic galaxy surveys. We then focus on the enhancement of the matter power spectrum in modified gravity (MG), the boost factor, using the Dvali-Gabadadze-Porrati (DGP) theory as a test case but developing a general approach that can be applied to other MG theories. After identifying the minimal simulation requirements for accurate DGP boost factors, we design and produce a COLA simulation suite that we use to train a neural network emulator for the DGP boost factor. Using \texttt{MG-AREPO} simulations as a reference, we estimate the emulator accuracy to be of $\sim 3\%$ up to $k=5\hompc$ at $0 \leq z \leq 2$. We make the emulator publicly available at: \url{https://github.com/BartolomeoF/nDGPemu}.} 

\begin{document}
\maketitle
\flushbottom

\section{Introduction}
\label{sec:Introduction}
% non-linear regime of structure formation for next-generation surveys
With the successful launch of the Euclid\footnote{\url{https://www.euclid-ec.org/}} satellite and the DESI\footnote{\url{https://www.desi.lbl.gov}} collaboration releasing their $1\%$ survey data \cite{DESI:2023dwi}, cosmology is living a very exciting time. The extraordinarily large amount of galaxies that these surveys will observe is fuelling the efforts of the scientific community aimed at accurately modelling the non-linear regime of structure formation to obtain precise cosmological constraints. 

% Modified gravity 
One of the key objectives of next-generation surveys is to shed light on the nature of gravity by constraining alternative gravity models. 
The simplest modified gravity (MG) theories can be described by means of a fifth force mediated by an additional scalar field. Given the stringent constraints coming from solar system tests of gravity, MG theories of interest to cosmology incorporate screening mechanisms which suppress the fifth force in specific environments.  Hence, the large-scale structure (LSS) of the Universe provides a unique laboratory to test the validity of GR and explore alternative gravity models. By comparing LSS probes with theoretical predictions, we can place constraints on the deviations from GR and potentially uncover new physics.

% role of cosmological simulations for accessing the non-linear regime
Yet, accessing the non-linear regime of structure formation presents significant challenges. Analytical techniques, such as perturbation theory, break down in this regime, and numerical simulations become essential. Cosmological simulations play a vital role in studying the formation and evolution of structures, enabling us to investigate the complex interplay between gravity, dark matter (DM), and baryonic physics. To simulate structure formation on the smallest scales, DM-only \nbody{} simulations are not enough and it is necessary to run hydrodynamical simulations which incorporate the effects of baryonic matter. However, despite the advances in simulating baryonic physics, hydrodynamical simulations show important discrepancies in their small-scales predictions \cite{Chisari:2019tus}, so weak lensing data below baryonic-feedback scales is discarded in cosmological analysis (e.g. see \cite{DES:2021wwk, Lange:2023khv}).

% role of COLA method to speed up cosmological simulations
Furthermore, cosmological simulations are computationally demanding, requiring substantial computational resources and time. To overcome these computational challenges, the development of efficient methods to speed up simulations has been a topic of active research. One such method is the COmoving Lagrangian Acceleration (COLA) method \cite{Tassev:2013pn}, which combines second-order Lagrangian Perturbation Theory (2LPT) with the Particle Mesh \nbody{} technique to accurately reproduce the non-linear dynamics of structure formation while significantly reducing computational costs. In particular, COLA exploits the fact that long-range forces perceived by the \nbody{} particles in a frame comoving with the 2LPT solution are weaker than the ones perceived in the Eulerian frame. This allows COLA simulations to use a coarser time-resolution while providing an accurate description of the matter distribution on large scales \cite{Tassev:2013pn,Izard:2015dja}. An interesting alternative approach recently developed in \cite{List:2023jxz}, proposes the use of non-symplectic integrators inspired by LPT to recover the correct large-scale evolution in PM simulations without the need to explicitly compute the 2LPT solution.

% previous important results regarding COLA simulations
The potential of the COLA method was already exploited in \cite{Koda:2015mca} to produce mock galaxy catalogues for the analysis of the WiggleZ survey \cite{Drinkwater:2009sd}, where it was shown that COLA simulations can be used to create synthetic galaxy catalogues with comparable accuracy to the full \nbody{} code Gadget \cite{gadget2} on scales $k \leq 0.2 \hompc$. Since then, the COLA method has seen several extensions:
\begin{itemize}
    \item to create more accurate halo catalogues \cite{Izard:2015dja},
    \item to produce matter distribution and mock halo catalogues on the light-cone \cite{Howlett:2015hfa, Izard:2017kma},
    \item to extend to MG theories the evolution of the matter distribution \cite{Valogiannis:2016ane,Winther:2017jof, Wright:2022krq} and the creation of mock galaxy catalogues \cite{Fiorini:2021dzs},
    \item to include the effects of massive neutrinos \cite{Wright:2017dkw}.
\end{itemize} 
Of particular interest to this work is the extension to MG theories presented in \cite{Winther:2017jof} and as implemented in the publicly available library \codeword{FML}\footnote{\url{https://fml.wintherscoming.no}}. This relies on the extension to MG of the 2LPT solutions and to an approximation of the screening mechanisms based on the analytic solution in spherical symmetry \cite{Winther:2014cia} which is provided for the Vainshtein and chameleon screening mechanisms. Recently, an alternative technique to approximate the Vainshtein screening \cite{Scoccimarro:2009eu} has been revised and implemented in a custom version of \codeword{FML-COLA} \cite{Brando:2023fzu}.

%  use of emulators in cosmology
Despite the speed-up brought by fast simulation methods like COLA, the computational cost of simulations is still too high to allow a fine sampling of the vast theory parameter space that is needed in Monte Carlo searches to infer precise parameter constraints. One possible solution to this problem is to approximate the predictions of cosmological simulations for specific summary statistics making use of machine learning techniques ~\cite{agarwal2012,habib2007} or fitting functions (see for instance \cite{Smith:2002dz}). The resulting approximating functions are normally referred to as emulators in the field and their key feature is that they provide theoretical predictions at extremely small computational cost. Accurate emulators for the matter power spectrum, one of the most elementary summary statistics that can be predicted thanks to simulations, are available for standard cosmologies \cite{Euclid:2020rfv, Angulo:2020vky}. Matter power spectrum emulators have been developed also in MG gravity theories such as  \cite{Winther:2019mus,Ramachandra:2020lue,Arnold:2021xtm, Saez-Casares:2023olw, Mauland:2023pjt} in Hu-Sawicki $f(R)$ model~\cite{Hu:2007nk}, and \cite{Harnois-Deraps:2022bie} in the Dvali-Gabadadze-Porrati (DGP) model~\cite{Dvali:2000hr}. Interesting alternative approaches based on the halo model were also developed to predict the matter power spectrum in MG theories \cite{Cataneo:2018cic, Bose:2022vwi, Gupta:2023rbf}.

Due to the many subtleties that characterise \nbody{} simulations and to the approximate nature of emulators, these tools need to be validated. This can be done for example by means of comparisons between different approaches \cite{Winther:2015wla, Euclid:2020rfv, Angulo:2020vky}. In the case of \nbody{} simulations, it is also important to check that the agreement with the other codes is not just due to a coincidence and that the predictions converge to the true answer when increasing resolution.

% statement of aim
In this work we push the COLA method to its limits to study the convergence of COLA predictions for the matter power spectrum and its enhancement in MG, the boost factor. We employ the results of this analysis to design and produce an emulator for the matter power spectrum in DGP gravity that we publicly release together with this paper. While in this work we focus on DGP gravity, the framework that we develop is quite general and can be applied to other MG theories. 

% overview of the paper
The remainder of this paper is organised as follows. In section~\ref{sec:Convergence} we provide an in-depth analysis of the convergence and validation of COLA simulations for prediction of the DM power spectrum in GR. In section~\ref{sec:Boost} we extend the convergence analysis to the nDGP boost factor for the DM power spectrum and present an emulator for this signal trained on COLA simulations. Finally, we come to our conclusions in section~\ref{sec:Conclusions}.

\section{Convergence tests}
\label{sec:Convergence}

% How to validate simulations
Cosmological simulations are subject to errors due to the finite resolution characterising their numerical implementation. In the case of {\it N}-body simulations, time-resolution, force-resolution and mass-resolution are important quantities which control the simulation accuracy \cite{Klypin:2017iwu,DeRose:2018xdj,Euclid:2020rfv}. To validate simulation results it is necessary to compare the output of simulations against well-established results like analytic predictions or previous simulations already validated (e.g., in \cite{Fiorini:2021dzs,Fiorini:2022srj} we validated COLA simulations by comparing them with ELEPHANT simulations). Furthermore, the self-consistency of a simulation technique can be established by performing relative convergence tests, where simulations with different resolutions are compared to understand the reliability of specific simulation settings. 

% What do we expect from increasing the resolution
The COLA method exploits the 2LPT and the Particle-Mesh (PM) method to evolve the DM distribution. The use of 2LPT allows us to reduce the time resolution of simulations without losing accuracy on large scales \cite{Izard:2015dja}. Conversely, increasing the time resolution makes the use of 2LPT less important and with sufficient time resolution, the simulations evolve like PM-only simulations. In this sense, we will refer to COLA simulations with high time resolution as PM simulations in the following sections. 
This equivalence is important since the accuracy of PM simulations (and therefore of COLA simulations in $\Lambda$CDM) is only limited by resolution and they converge to full {\it N}-body simulations results with increasing resolution \cite{Hernandez-Aguayo:2021kuh}. 
However, when comparing COLA with PM-only algorithms, it is not fair to compare their computational cost using the same number of time-steps as done in \cite{Klypin:2017iwu}. In fact, COLA does not necessarily need as many time-steps as PM-only simulations to achieve comparable accuracy. This is evident in the low time-resolution case, where the 2LPT guarantees that COLA provides accurate results on large scales while PM-only simulations already show significant deviations \cite{Izard:2015dja}. Even in the high time-resolution case, COLA can achieve the same accuracy as PM-only codes with fewer time-steps. Indeed, the latter needs a finer time-stepping at high redshift to deal with the rapidly changing growth factor \cite{Klypin:2017iwu} while the 2LPT guarantees good accuracy at high redshift in COLA even with larger time-steps \cite{Tassev:2013pn,Izard:2015dja}. For instance, the PM code \codeword{GLAM} needs 147 time-steps to have $\Delta a / a \approx 0.014$ at $z=0$ \cite{Klypin:2017iwu} while COLA needs only $\approx70$ time-steps for the same low-redshift time resolution.

% section overview
The focus of this section is to validate COLA simulations for the predictions of the matter power spectrum. To do so, we first introduce the cosmological emulators in subsection~\ref{ssec:CosmoEmu}, then we compare COLA results with full {\it N}-body simulations and cosmological emulators in subsection~\ref{ssec:COLAvsCosmoEmuandArepo}, and finally, we study the relative-convergence of COLA simulations for the matter power spectrum in subsection~\ref{ssec:Convergence}.

\subsection{Cosmological emulators}
\label{ssec:CosmoEmu}
Cosmological emulators interpolate the results of computationally expensive cosmological simulations in high-dimensional parameter spaces. Two notable such emulators are the Bacco emulator~\cite{Angulo:2020vky} and the Euclid Emulator 2 (EE2)~\cite{Euclid:2020rfv}. While these emulators differ on many aspects (e.g. the interpolation techniques employed, the parameter space spanned and the \nbody{} techniques used for the training set) they have been shown to be in agreement with each other within the claimed accuracy \cite{Euclid:2020rfv,Brando:2022gvg} for several test cosmologies. 
  
\begin{figure}
\centering 
\includegraphics[width=.85\textwidth]{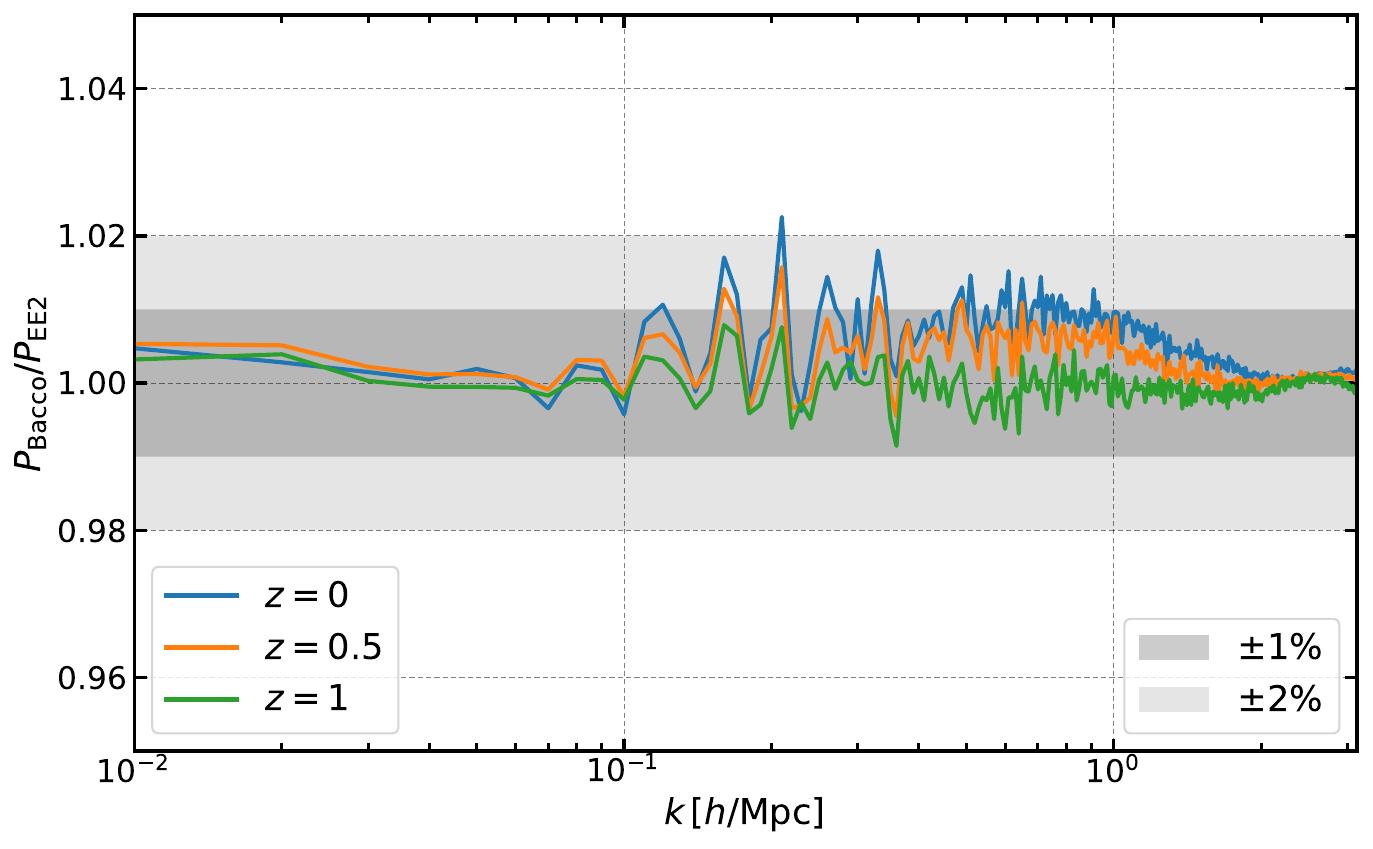}
\caption{\label{Fig:Pk_BaccoVsEE2} Ratio of Bacco and EE2 power spectra for the cosmology defined in eq~\eqref{ArepoCosmo}.}
\end{figure}

Here in particular we are interested in the agreement between these two emulators for a specific cosmology, the one that was used to run the {\it N}-body simulation suite that we will discuss in subsection~\ref{ssec:COLAvsCosmoEmuandArepo}.  In figure~\ref{Fig:Pk_BaccoVsEE2} we show the ratio of the matter power spectra estimated with Bacco and EE2 at redshifts $0$, $0.5$ and $1$ for the above-mentioned cosmology (see eq~\ref{ArepoCosmo}). The agreement is within $2\%$ at all the scales considered. At $z=1$, the agreement is well within $1\%$. This is consistent with the accuracy claimed by the two emulators and gives additional reliability to the emulators' results, in particular at redshift $z=1$, that we use in the next subsection to prove the accuracy of COLA simulations for predictions of the matter power spectrum. 

\subsection{Comparison of COLA with emulators and {\it N}-body simulations}
\label{ssec:COLAvsCosmoEmuandArepo}
As reference results to compare COLA with, we make use of the power spectra obtained from a suite of DM-only simulations in GR and nDGP gravity\footnote{The nDGP results (for $H_0 r_c$ values of 0.5, 1, 2, 5) are used in section~\ref{sec:Boost}.} ran with \codeword{MG-AREPO}  ~\cite{Arepo_fR,Arepo_nDGP} which uses the tree-particle-mesh technique for gravitational interactions and a moving mesh to solve the hydrodynamic equations (see references~\cite{Arepo, gadget2} for more on the \nbody{} techniques at the base of this code). The simulations were started at redshift $z=127$ using the cosmological parameters
% Cosmology
\begin{equation}
\begin{array}{ccc}
\Omega_{m,0}=0.3089 \, , & \Omega_{\Lambda,0}=0.6911 \, , & \Omega_{b,0}=0.0486 \, , \\
n_{s}=0.9667 \, , & \sigma_{8}=0.8159 \, , & h=0.6774 \, ,
\end{array}
\label{ArepoCosmo}
\end{equation}
and evolved down to redshift $z=0$ using $1024^3$ particles in a $(1{\rm Gpc}/h)^3$ box~\cite{Mitchell:2021aex}.

To assess the accuracy of COLA simulations we produce a suite of simulations in a $L = 512$Mpc$/h$ box, with the same cosmology of the Arepo simulations and with different mass resolutions ($\NP=512, 1024, 2048$), time resolutions ($\NS=25\times[1, 2, 3]= 25, 50, 75$) and with force resolutions depending on the mass resolutions ($\NM=[1,2,3,4]\times\NP$ for $\NP=512,1024$ and $\NM=[1,1.5,2]\times\NP$ for $\NP=2048$). We start the simulations at redshift $z=19$ and stop them at redshift $z=1$. The choice of 25 time-steps as the low time-resolution case is motivated by the time-step size of typical COLA configurations ($\Delta a \approx 0.02$). In the high time-resolution case, where the late-time evolution is PM-like, 75 time-steps produce a time-step size of $\Delta a \approx 0.006$. This is the same step size of the reference simulations used in the convergence tests of \codeword{GLAM} at $z<3$ \cite{Klypin:2017iwu}.

\begin{figure}[t]
\centering 
\includegraphics[width=.85\textwidth]{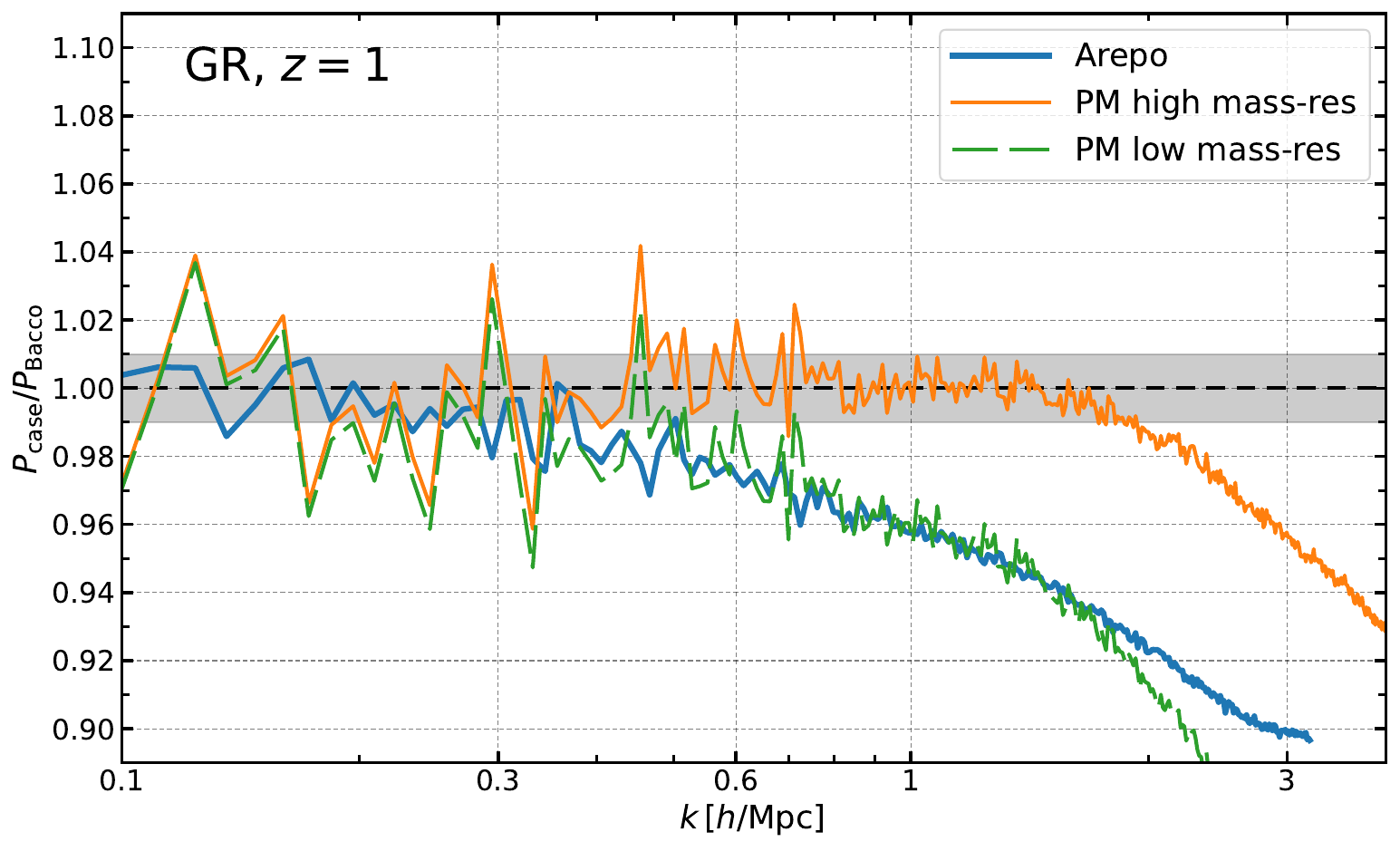}
\caption{\label{Fig:COLA_Arepo_over_Bacco} Comparison of PM simulations matter power spectra with Bacco and Arepo power spectra. The high mass-resolution PM run agrees with Bacco up to $k\sim2h/$Mpc within $1\%$. Arepo does not agree with Bacco but is in better agreement with the low mass-resolution PM run which has approximately the same mass resolution as the Arepo simulation.}
\end{figure}

 In the next subsection, we will use the highest resolution PM run ($\NP=2048$, $\NM=4096$, $\NS=75$) as the reference to assess the accuracy of all the other runs. Before that, we validate the highest resolution run by comparing its matter power spectrum with the ones from the Bacco emulator and from the Arepo simulations. In Figure~\ref{Fig:COLA_Arepo_over_Bacco} %
we plot the ratio of the matter power spectra of PM and Arepo simulations with the Bacco power spectrum for the cosmology of eq.~\eqref{ArepoCosmo}, in GR at redshift 1. The first thing that comes to the eye, is that the Arepo power spectrum (green line) departs from Bacco already at $k\sim 0.6h$/Mpc, and at $k\sim 1h/$Mpc the discrepancy is already at $4\%$. We attribute this to the low mass resolutions of these Arepo simulations ($M_{\rm part} \approx8\cdot 10^{10} \Msun$) \cite{Euclid:2020rfv}, and to motivate this claim we add to the comparison the PM simulation for $\NP=512$, $\NM=2048$, $\NS=75$ (blue line) which has a similar mass resolution of Arepo simulations. This is in good agreement with Arepo up to  $k\sim 2h/$Mpc. The blue line instead shows that the higher resolution PM run is in sub-percent agreement with Bacco up to $k\sim 2 \hompc$. 

\subsection{Relative convergence of COLA simulations}
\label{ssec:Convergence}
\begin{figure}[t]
\centering 
\includegraphics[width=.99\textwidth]{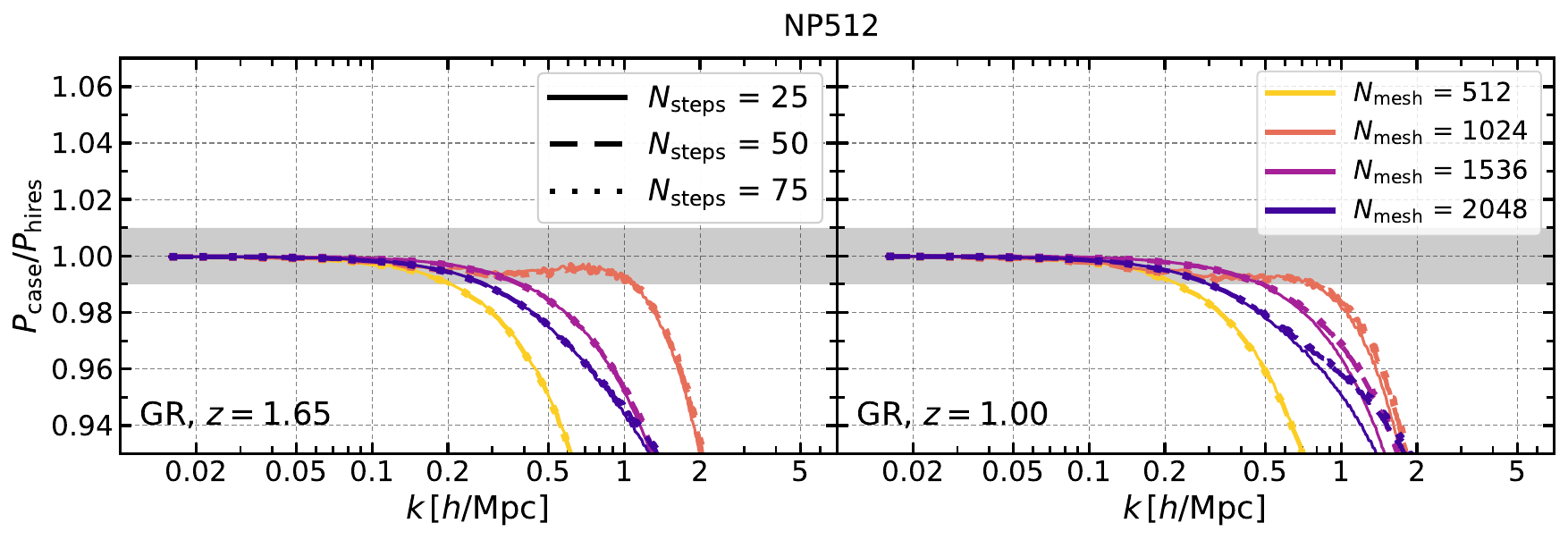}
\caption{\label{Fig:GR_NP512} Ratio of the matter power spectrum of simulations with 512 particles for the force and time resolutions shown in the legend with that of the high-resolution run.}
\end{figure}

\begin{figure}[t]
\centering 
\includegraphics[width=.99\textwidth]{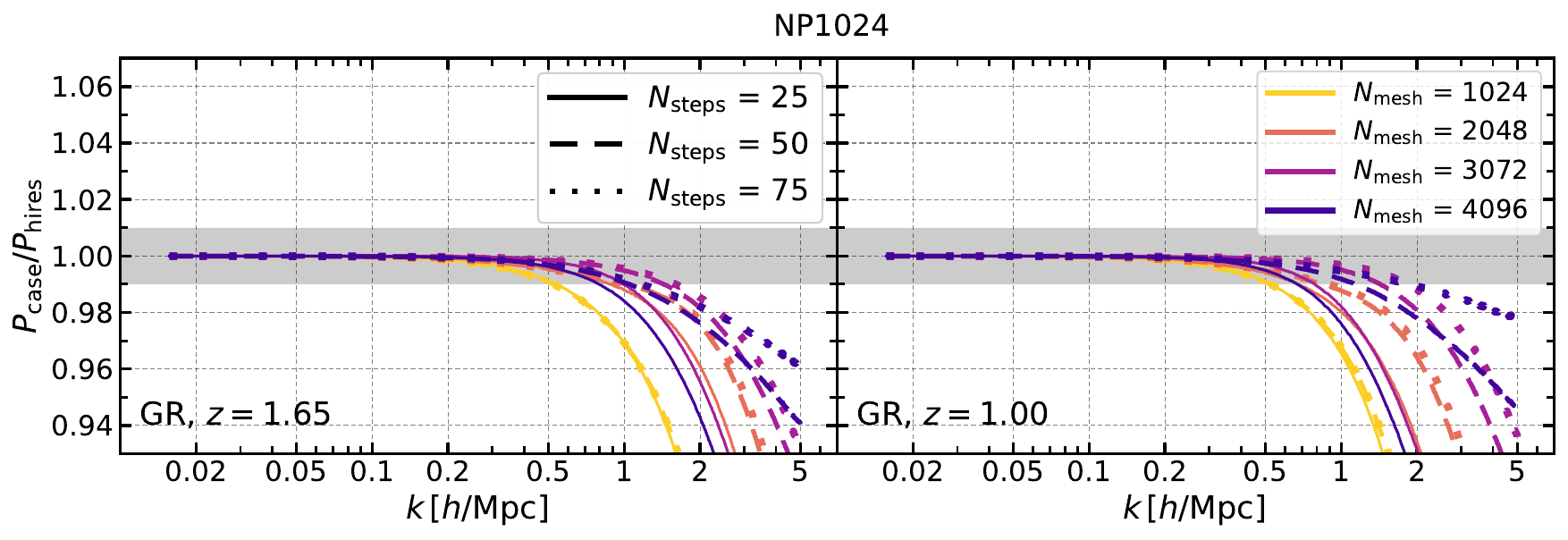}
\caption{\label{Fig:GR_NP1024} Ratio of the matter power spectrum of simulations with 1024 particles for the force and time resolutions shown in the legend with that of the high-resolution run.}
\end{figure}

\begin{figure}[t]
\centering 
\includegraphics[width=.99\textwidth]{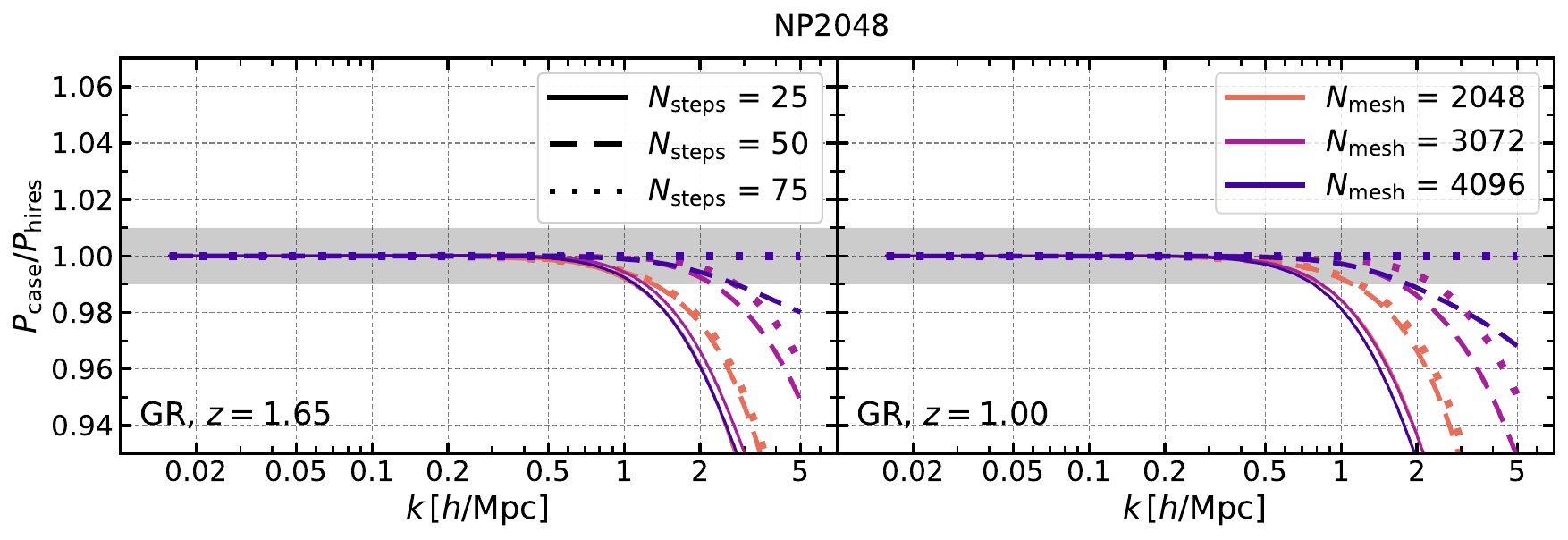}
\caption{\label{Fig:GR_NP2048} Ratio of the matter power spectrum of simulations with 2048 particles for the force and time resolutions shown in the legend with that of the high-resolution run.}
\end{figure}

% Show convergence towards the hi-res COLA run (the PM limit) and discuss the role of mass/force/time resolution in PM simulations and in particular in COLA.
With all the simulations of the convergence suite, we compare the accuracy of all the different settings using the high-resolution run as the reference.
To summarise the results, we show in different figures different mass resolutions, figure~\ref{Fig:GR_NP512} for $\NP=512$, figure~\ref{Fig:GR_NP1024} for $\NP=1024$, figure~\ref{Fig:GR_NP2048} for $\NP=2048$, and split each figure into 2 panels for two redshift values $z=1.65, 1$ respectively on the left and on the right%
\footnote{This choice of redshift values is motivated by the expected redshift localisation of H$\alpha$ emitters in the Euclid spectroscopic survey \cite{Euclid:2019clj}.}. In each panel, we show the ratio of the power spectra in the COLA run with that of the high-resolution run for the various force and time resolutions described in the legends .

For the $\NP=512$ case, we notice that the results are almost insensitive to the time resolution and do not converge towards the high-resolution result. This is attributed to the fact that the low mass resolution has a strong effect on the scales considered in the comparison, especially for higher redshifts. In addition to this, we notice that the convergence is non-monotonic: the fact that $\NM=1024$ results have significantly larger power than the ones of $\NM=1536$ and $\NM=2048$ is somehow surprising. From additional tests that we have run, we could identify this behaviour as being due to discreteness effects happening at high redshifts (see \cite{Michaux:2020yis,List:2023kbb} for recent progress on this topic) when the particles' displacements from the Fourier grid are small in combination with the continuous Fourier space kernels used for the force computation in the PM scheme\footnote{In the FML library it is now possible to choose the technique adopted by the PM solver for the force computation. In particular, it is possible to choose the same finite differencing scheme used in codes like FastPM~\cite{Feng:2016yqz} and GLAM \cite{Klypin:2017iwu} which seems to avoid these problems of non-monotonic convergence in low mass-resolution PM simulations.}. While this effect is moderate when starting the simulations at $z=19$, it becomes more important when starting the simulations at higher redshift.

For the $\NP=1024$ and $2048$ cases, the convergence towards the high-resolution PM result happens only if the force and time resolutions are increased accordingly. This condition is particularly evident in the simulations with $\NP = 1024$, where increasing the number of time steps from 25 to 75 keeping the fixed force resolution of $\NM = 1024$ does not improve the convergence toward the high-resolution run. Similarly, increasing the force resolution above $\NM = 2048$ but using only 25 time-steps does not improve the agreement with the highest-resolution run. We notice that, in the $\NP=1024$ case, the results for high force and time resolution appear to be more converged at low redshift than at high redshift: this is due to the fact that discreteness effects are more important at early times while force resolution becomes more important at late times.

\begin{figure}[t]
\centering 
\includegraphics[width=.85\textwidth]{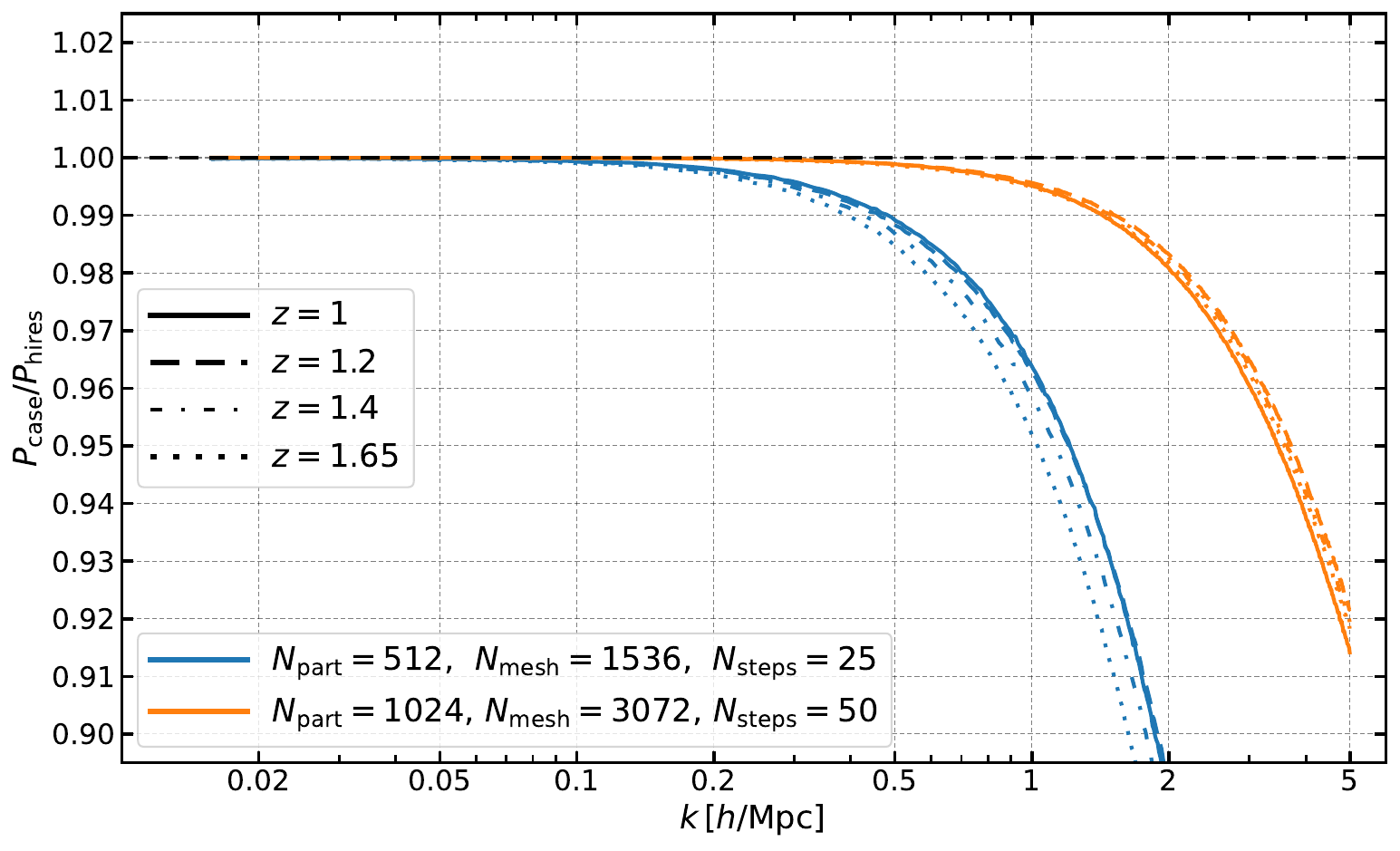}
\caption{\label{Fig:COLA_rel_convergence} Convergence of matter power spectra towards the high-resolution result as mass-, force- and time-resolution are simultaneously increased (as indicated by the colour legend). Different line styles are used to distinguish results at different redshift values as indicated in the line-style legend.}
\end{figure}

To assess the convergence of the highest-resolution run used as a reference in the previous figures, we compare in figure~\ref{Fig:COLA_rel_convergence} the power spectra of lower-resolution simulations with the highest-resolution one, as done previously, but now increasing mass, force and time resolution simultaneously. The agreement between the $\NP=1024$, $\NM=3072$, $\NS=50$ run and the highest-resolution run indicates that the latter is converged at sub-percent level up to $k\sim1\hompc$. This is in line with the result of the comparison with the Bacco emulator.

\section{Modified gravity boost factor}
\label{sec:Boost}

%===---###===---###===---###===---###===---###===---###===---###===---###===---###===---###===---###===---###===---###===---###===---###===---###===---###===---###===---###===---###
Another interesting application of COLA simulations is to predict the matter power spectrum in MG \cite{Winther:2019mus}. As non-linear matter power spectrum emulators are already available, it is possible to obtain the matter power spectrum in MG by multiplying the matter power spectrum in GR with the MG boost factor
\begin{equation}
    B_{\rm MG}(k,z) \equiv \frac{P_{\rm MG}(k,z)}{P_{\rm GR}(k,z)} \, . 
\end{equation}
In general, the MG boost factor depends on MG parameters as well as on cosmological parameters and redshift. 
In the following, we focus on the DGP model and use COLA simulations to train an emulator for the matter power spectrum in this gravity theory. Nonetheless, the method that we develop is not specific to the DGP model and can be used to produce emulators for other MG theories.

The DGP gravity is a braneworld model in 5-dimensions, which has been thoroughly studied by the modified gravity community due to its interesting cosmological features \cite{Deffayet:2000uy, Luty:2003vm,Koyama:2007ih}. This theory has 2 branches of solutions for the Hubble expansion on the brane~\cite{Deffayet:2000uy}: a self-accelerating branch with ghost instabilities on a de Sitter background~\cite{Luty:2003vm, Koyama:2007za}, and a normal branch which requires fine-tuning in the form of an effective dark energy to recover the observed $\Lambda$CDM-like expansion history \cite{Schmidt:2009sg, Lombriser:2009xg}. Nonetheless, the normal branch of the DGP theory (nDGP) also presents a non-trivial fifth force dynamics which can be described by the Vainshtein mechanism \cite{Deffayet:2001uk}, so this theory was often used as a test-case representative for the class of Vainshtein-screened theories. In particular, several \nbody{} codes were developed to simulate non-linear structure formation in nDGP theory (e.g. \texttt{DGPM}~\cite{Schmidt:2009sg}, \texttt{ECOSMOG}~\cite{Li:2011vk}, \texttt{MG-AREPO}~\cite{Hernandez-Aguayo:2020kgq}, \texttt{MG-GLAM}~\cite{Hernandez-Aguayo:2021kuh}) and two of them, \texttt{DGPM} and \texttt{ECOSMOG}, were showed to give consistent prediction for the matter power spectrum boost-factor in \cite{winther15}.

To develop the nDGP emulator, we start by studying the simulation requirements to accurately predict the nDGP boost factor with COLA and investigate the sensitivity of the boost factor to the theory parameters in subsection~\ref{Sec:AccSens}. Then we present the simulations suite and the data processing used to create the emulator's data sets in subsection~\ref{ssec:SimsAndData} and finally, we discuss the emulator's training and performance in subsection~\ref{Sec:Emulator}.

\subsection{Accuracy and sensitivity}
\label{Sec:AccSens}
The computational cost of cosmological simulations depends on several parameters. In particular, it is very sensitive to volume and resolution. With this in mind, we want to find the optimal simulation settings that allow us to have the target accuracy while probing large enough scales but without requiring excessive computational resources.

To define the simulation requirements for the nDGP emulator, we start by assessing the accuracy of COLA in predicting the boost factor of nDGP theories against Arepo simulations. To this extent, we run high-resolution PM simulations\footnote{We emphasize that PM here refers to the high time-resolution used by these COLA simulations at late time, which makes the 2LPT contribution important only at early times as discussed in section~\ref{sec:Convergence}. 
It is important to note that the screening mechanism is implemented only approximately in COLA simulations for DGP gravity ~\cite{Winther:2014cia, Winther:2017jof, Fiorini:2021dzs, Fiorini:2023uzw}. For an alternative screening approximation for general Vainshtein mechanism see~\cite{Brando:2023fzu}. On the other hand, Arepo simulations solve the non-linear scalar field exactly.} with the same cosmology of Arepo simulations for several values of the nDGP parameter, $H_0 r_c$. We simulate the evolution of $2048^3$ particles in a box of side $L = 1024 \mpcoh$ using $6048^3$ mesh-grids and 75 time-steps from redshift $z_{\rm ini}=127$ to $z_{\rm fin}=1$. The mass resolutions that we use in the PM simulations is $M_{\rm part} \sim 1 \cdot 10^{10} \Msun$, roughly 8 times higher than the one of the Arepo simulations. The force resolution is $\LF \equiv L/\NM \approx 0.17 \mpcoh$ and the time resolution is $\Delta a \approx 0.006$. We run the simulations in GR and nDGP from the same IC. For nDGP simulations we use the 4 values of the $H_0 r_c$ parameter ($H_0 r_c= $0.5, 1, 2, 5) used in the Arepo simulations suite.

\begin{figure}
\centering
\includegraphics[width=.8\textwidth]{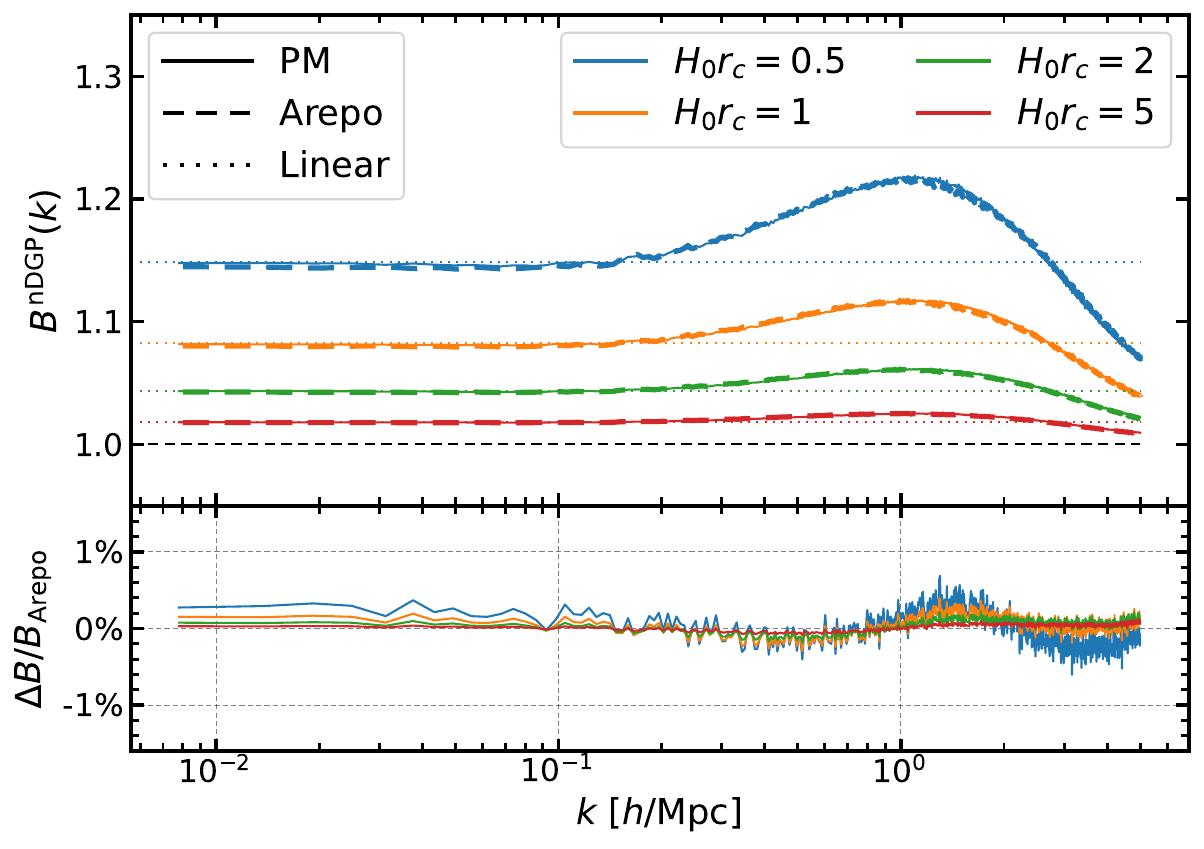}
\caption{\label{fig:BnDGP_COLAvsArepo} Ratio of the nDGP boost factors at redshift $z=1$ between PM simulations and Arepo simulations in 4 nDGP gravity models as indicated in the legend.}
\end{figure}

Figure~\ref{fig:BnDGP_COLAvsArepo} shows a comparison between the nDGP boost factors of PM and Arepo simulations for the different values of the $H_0 r_c$ parameter. We also include in the comparison the nDGP boost factor estimated from linear theory. In the top panel, we show the nDGP boost factors, while in the bottom we show the relative difference between the PM and linear boost factors from Arepo boost factors. On large scales, PM results seem to be more consistent with linear theory than the ones from Arepo simulations. Overall, we find $\lesssim 0.5 \%$ agreement in all gravity models up to $k \sim 5 \hompc$ between the high-resolution simulations and Arepo simulations. This means that the screening approximation used in COLA simulations in nDGP theory~\cite{Winther:2017jof, Fiorini:2021dzs}, once tuned on full {\it N}-body simulations for a single value of $H_0 r_c$, is sub-percent accurate in reproducing the power spectrum boost factor of nDGP theories for a wide range of the parameter $H_0 r_c$ and for different redshift values \footnote{We are using here the same screening settings adopted in \cite{Fiorini:2021dzs} and tuned against Ecosmog~\cite{Li:2011vk} results at $z\sim0.5$.}. 

\begin{figure}
\centering
\includegraphics[width=.98\textwidth]{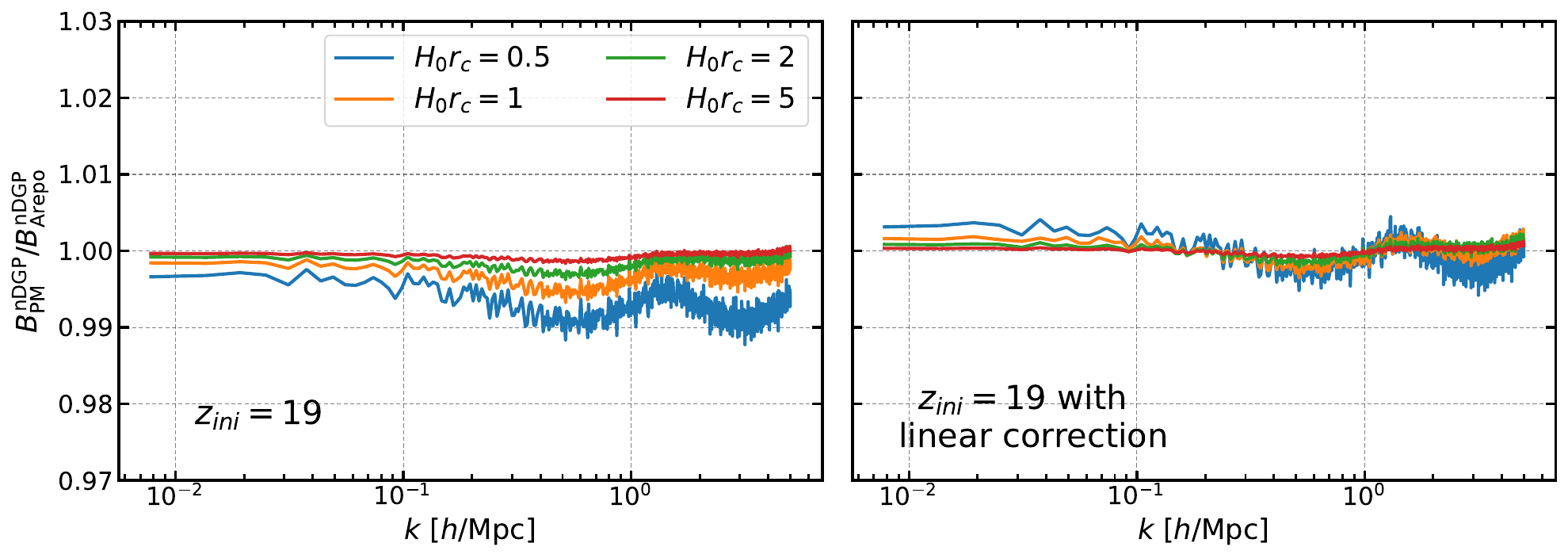}
\caption{\label{fig:BnDGP_COLAz19LinCorr} Ratio of the nDGP boost factors at redshift $z=1$ between PM simulations started at redshift 19 and Arepo simulations in 4 nDGP gravity models as indicated in the legend. The left panel shows the raw results from simulations, and the right panel shows the results after applying the linear correction in eq~\eqref{GrowthLinCorr}.}
\end{figure}

The COLA simulations that we run in nDGP use IC produced with the back-scaling approach. The IC are set in such a way that the IC power spectrum is the same for nDGP and GR simulations. This is done by integrating the growth factors from matter domination at redshift $z=499$ and by normalising them at the redshift of IC $z_{\rm ini}$
\begin{equation}
    D^{z_{\rm ini}} (z_{\rm ini}) = 1 \, .
\end{equation}
Due to this choice of normalisation, COLA simulations ignore the fact that nDGP can be different from GR also at higher redshift. In particular, COLA simulations are usually started at redshift $z_{\rm ini}=19$, while nDGP effects can be significant up to redshift $z \sim 100$. We can correct this by multiplying the final power spectrum by%
\footnote{For the sake of generality, we stress that this correction should not be used when providing input power spectra for the MG simulations which already take into account MG effects at high redshift. Furthermore, this correction may be irrelevant for MG models whose modifications are weaker than those of the nDGP model at high redshift.}%
\begin{equation}\label{GrowthLinCorr}
    \mathcal{A}_{\rm lin} (z) = \left(\frac{D_{\rm nDGP}^{z_{\rm ini}=127}(z) \cdot D_{\rm GR}^{z_{\rm ini}=19}(z) }{D_{\rm GR}^{z_{\rm ini}=127}(z) \cdot D_{\rm nDGP}^{z_{\rm ini}=19}(z)} \right)^2 \, ,
\end{equation}
which has the effect of re-scaling the amplitude of the power spectrum as if the simulations were started at redshift $z=127$. This approach neglects the effects of non-linearity before $z_{\rm ini}=19$, but these are sub-dominant due to the weakness of the nDGP fifth force at early times. The importance of this correction is evident from figure~\ref{fig:BnDGP_COLAz19LinCorr}, where we show that the agreement with Arepo of the nDGP boost factors from PM simulations started at $z_{\rm ini}=19$ is improved when applying the linear correction. In fact, the raw results of $z_{\rm ini}=19$ PM simulations show $1\%$ deviations from Arepo (left panel), while they are in better than $0.5\%$ agreement with Arepo once corrected using eq~\eqref{GrowthLinCorr} (right panel).

\begin{figure}[t]
\centering
\includegraphics[width=.98\textwidth]{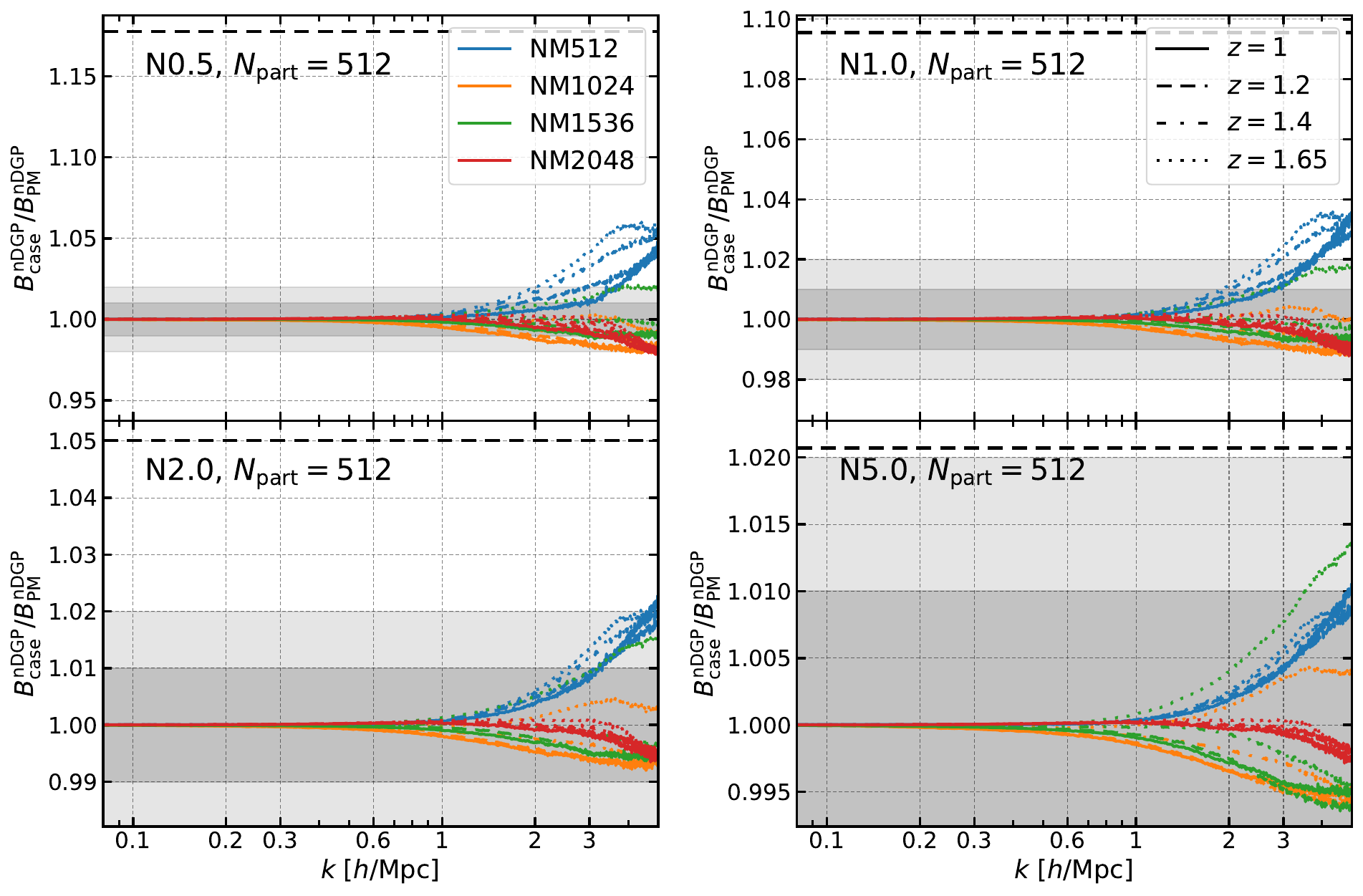}
\caption{\label{fig:BnDGP_LowVsHires} Ratio of the nDGP boost factors in low mass-resolution COLA simulations with that in high-resolution PM simulations for different force resolutions, $\NM = 512$, $\NM = 1024$, $\NM = 1536$ and $\NM = 2048$. In different panels, we show results for different nDGP gravity models as indicated in the top-left corner of each panel. The black dashed lines show the typical amplitude of the boost factor in the different gravity models for visual reference.}
\end{figure}

Having tested the accuracy of COLA simulations in the PM-limit in reproducing the nDGP boost factor, we want to test the impact of force and mass resolution on the nDGP boost factor in COLA simulations to find the optimal setup to span the theory parameter space running a large number of simulations. To do so, we run more COLA simulations in a box of side $L=512\hompc$ with lower mass-resolution ($M_{\rm part} \sim 8 \cdot 10^{10} \Msun$, similarly to Arepo simulations), with time-resolution $\Delta a \approx 0.02$ and with 4 force resolutions: $\NM=512$ (low force resolution), $\NM=1024$, $\NM=1536$,  $\NM=2048$ (high force resolution). We compare the results for the nDGP boost factor of low mass-resolutions COLA simulations with high-resolutions PM simulations in figure~\ref{fig:BnDGP_LowVsHires} where we can see that the results are converged at sub-percent level up to $k\sim 1 \hompc$ and the force resolution has a significant impact on the boost factor only on very small scales, $k> 2 \hompc$. The results of the intermediate cases, $\NM=1024$ and $\NM=1536$, are consistent with the high-resolution PM simulations within $2\%$ up to $k\sim 5 \hompc$. %, while the mass resolution has a negligible impact on the nDGP boost factor at all scales.

Before choosing a strategy to sample the parameter space for producing the emulator's training set, we explore the impact that each parameter has on the boost factor. To do so we first investigate the range of values that the parameter $H_0 r_c$ should span: for stronger deviations from GR we choose $\sim 50\%$ deviations as the upper limit, while for the other limit, we want to "smoothly" connect to GR so we look for models where the deviation is  $< 1\%$ (i.e., below our target accuracy). Based on previous results in the literature \cite{Ruan:2021wup}, we select $H_0 r_c = 0.2$ as the strong deviation case and $H_0 r_c = 20$ as the weak deviation case.

\begin{figure}
\centering
\includegraphics[width=.98\textwidth]{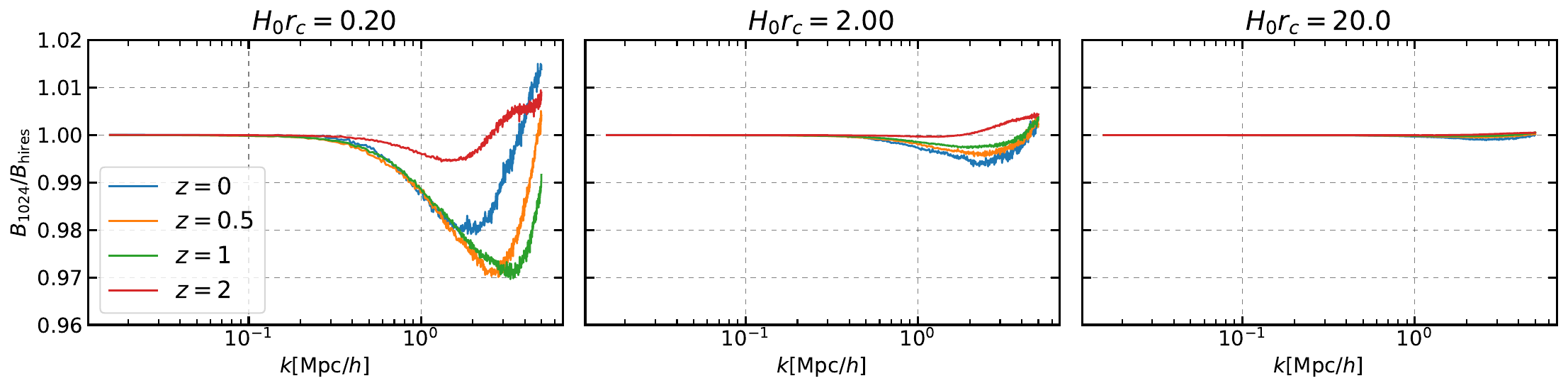}
\caption{\label{fig:nDGP_BoostConv_WideRange} Ratio of the nDGP boost factors obtained with low-resolution COLA simulations ($\NP = 512$, $\NM = 1024$) with respect to the same boost factors but obtained with high-resolution PM simulations ($\NP = 1024$, $\NM = 3072$). The legend describes the redshift value of each line. }
\end{figure}

The results of~\cite{Brando:2022gvg} show that the response function of the DM power spectrum for changes in cosmological parameters estimated with COLA agrees with cosmological emulators in a wide range of cosmologies when using $\LF \equiv L/\NP = 0.5 \mpcoh$. We shall use at least the same force resolution here for the boost factor since we want to predict it for a similar range of cosmologies. As an additional test to check the accuracy of the boost factor for the wider range of gravity parameters, we produce high-resolution PM simulations and use them to make a comparison with the boost factors of $\LF = 0.5 \mpcoh$ simulations.  %Having previously checked the accuracy of $N_{\rm mesh}= 3 N_{\rm part}$ simulations, we test the accuracy loss if using $N_{\rm mesh}= 2 N_{\rm part}$. 
Using a default box side of $512 \mpcoh$, we in fact compare the boost factors obtained using $512^3$ particles and $\NM= 1024$ with the ones obtained using $1024^3$ particles and $\NM= 3072$. We show the ratio of the two boost factors for 4 redshift values (legend) and 3 values of the parameter $H_0 r_c$ (plot titles) in figure~\ref{fig:nDGP_BoostConv_WideRange}. In the most extreme case, the difference in the two boost factors is $ \lesssim 1\%$ up to $k= 1 \hompc$ ($\lesssim 3\%$ up to $k= 5 \hompc$). This is in line with our target accuracy so we rely on these low-resolution settings for the simulations that we use to conduct the rest of our analysis. 

%===---###===---###===---###===---###===---###===---###===---###===---###===---###===---###===---###===---###===---###===---###===---###===---###===---###===---###===---###===---###
% \subsubsection{Parameter space}\label{Sec:ParSpa}

\begin{figure}[t]
\centering 
\includegraphics[width=.98\textwidth]{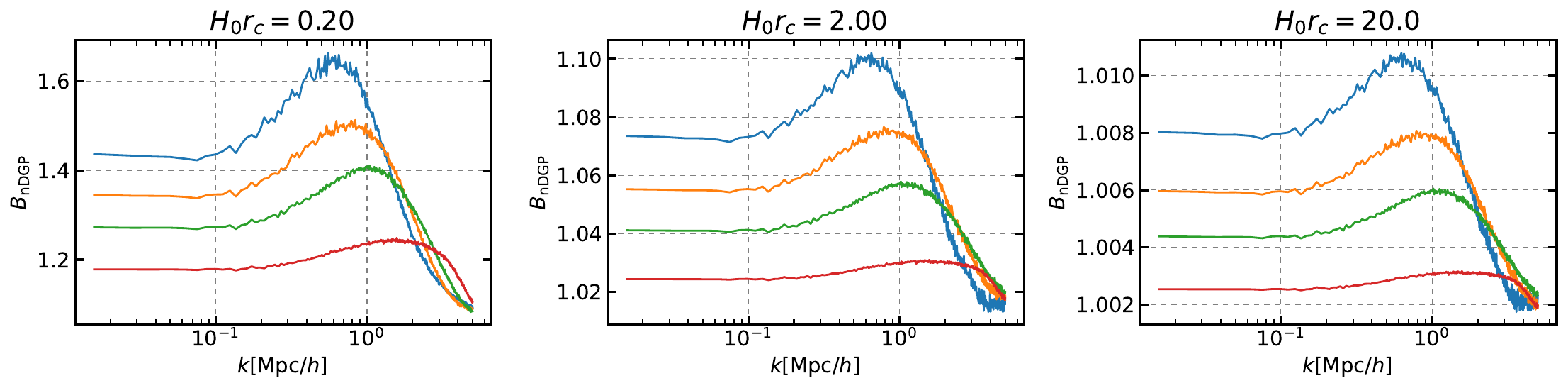}
\caption{\label{fig:Boost} nDGP boost factor for three values of the parameter $H_0 r_c$. The legend in figure~\ref{fig:nDGP_BoostConv_WideRange} describes the redshift value of each line. }
\end{figure}

\begin{figure}[t]
\centering
\includegraphics[width=.8\textwidth]{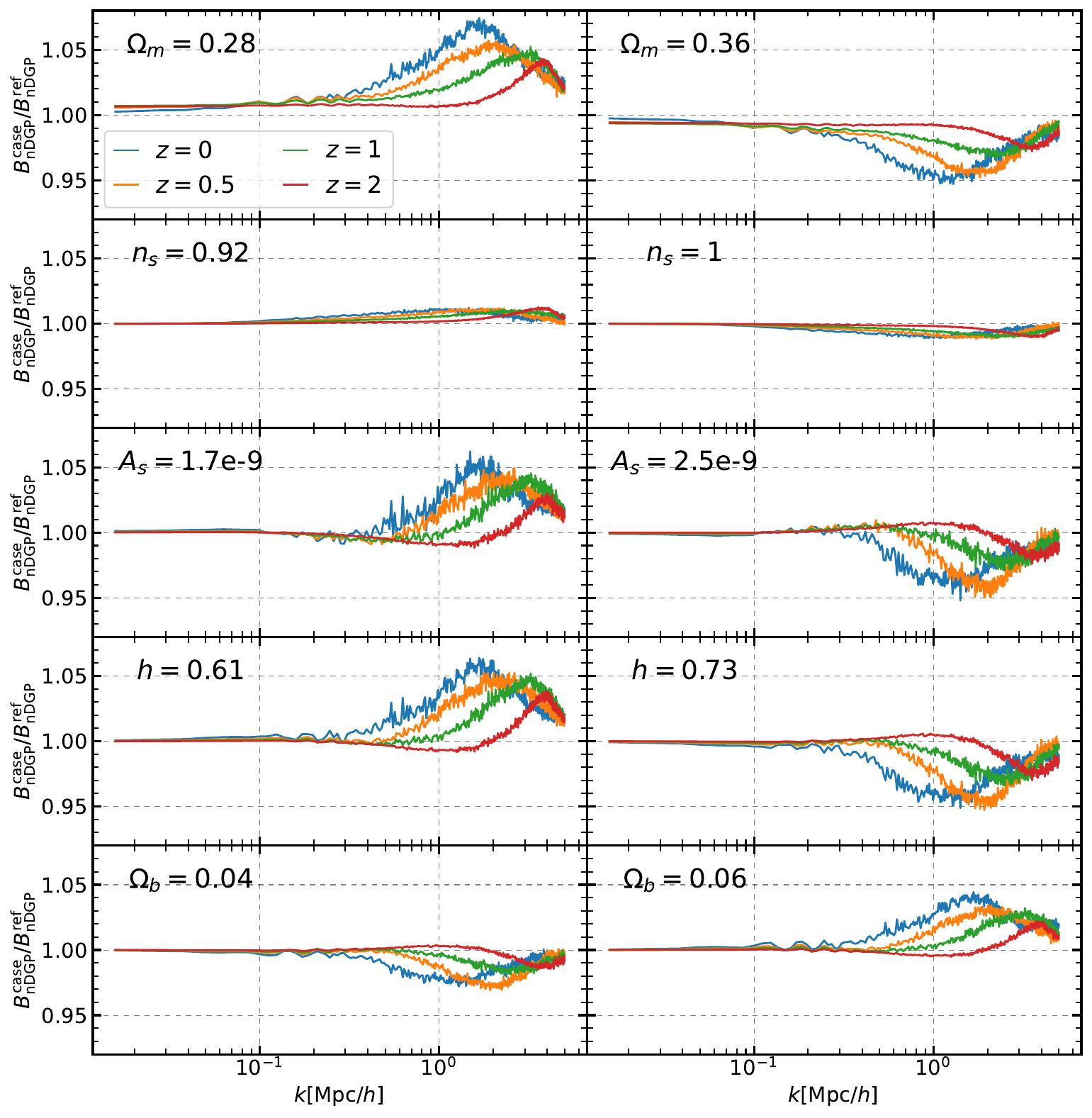}
\caption{\label{fig:CosmoEffect1} Change of the nDGP boost factor for $H_0 r_c = 0.2$ due to the variation of each cosmological parameter (specified in each panel) with respect to the reference cosmology. The redshift values of each line are as described in the legend.}
\end{figure}

\begin{figure}[t]
\centering
\includegraphics[width=.8\textwidth]{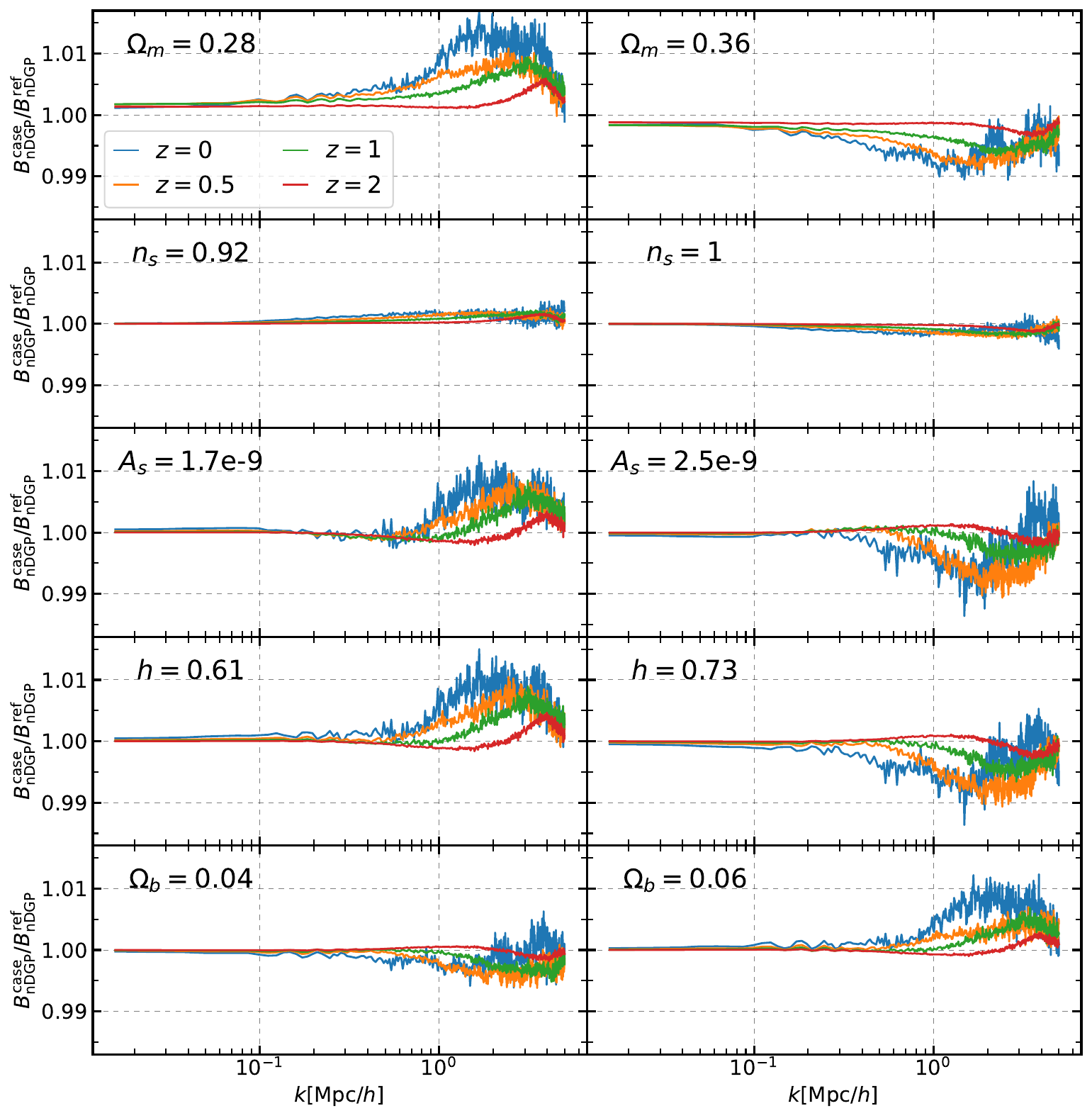}
\caption{\label{fig:CosmoEffect2} Same as figure~\ref{fig:CosmoEffect1} but for $H_0 r_c=2$}
\end{figure}

\begin{figure}[t]
\centering
\includegraphics[width=.8\textwidth]{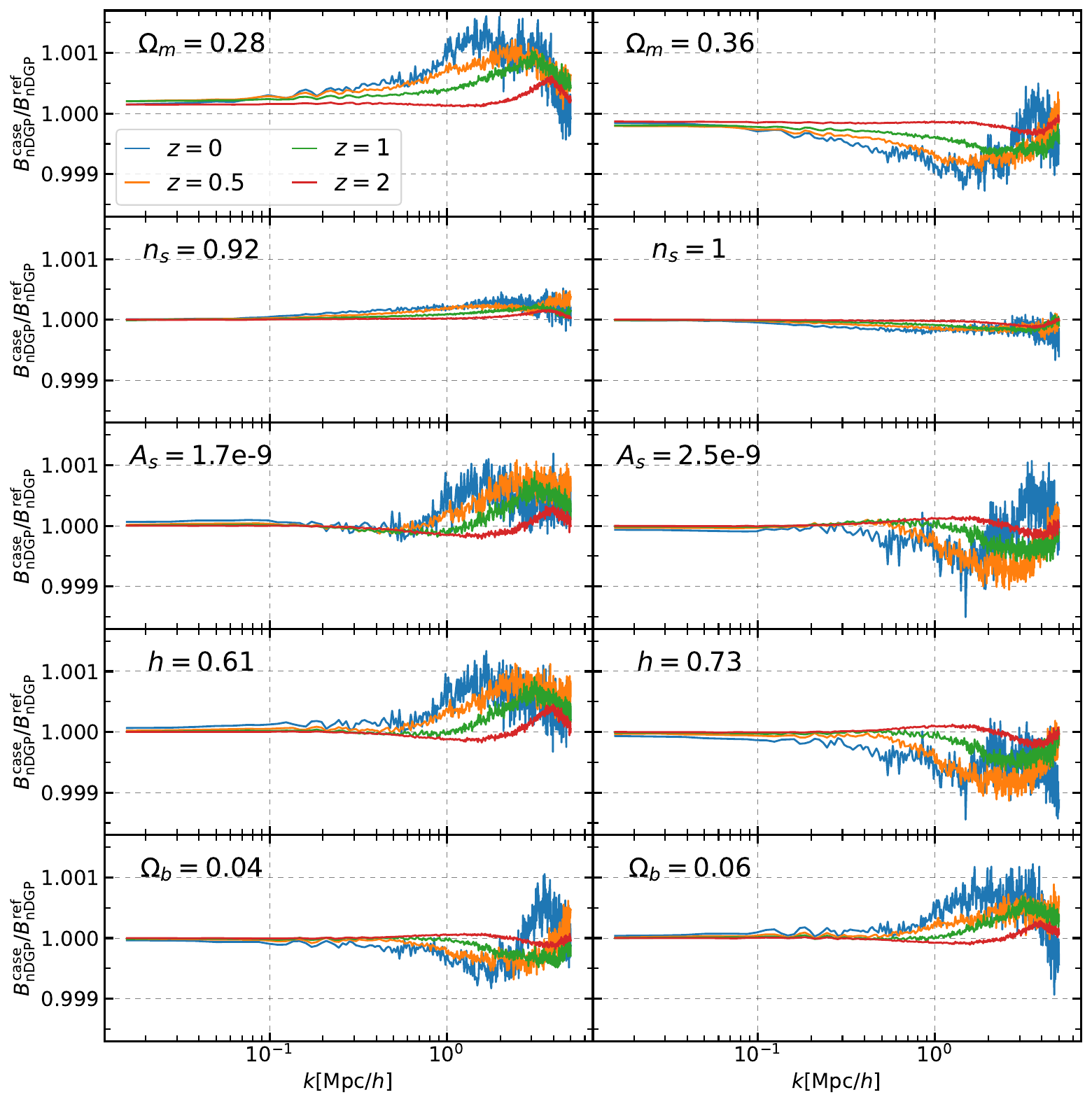}
\caption{\label{fig:CosmoEffect3} Same as figure~\ref{fig:CosmoEffect1} but for $H_0 r_c=20$}
\end{figure}

The boost factor of nDGP is strongly dependent on the theory parameter $H_0 r_c$ and on the redshift as shown in figure~\ref{fig:Boost}, where we plot the boost factors for three values of $H_0 r_c$ (titles) and 4 redshift values (legend). However, the cosmological dependence of the nDGP boost factor is quite weak. 
In support of this statement, we show the effect of changing each cosmological parameter on the nDGP boost factor for $H_0 r_c =0.2$ in figure~\ref{fig:CosmoEffect1}, $H_0 r_c =2$ in figure~\ref{fig:CosmoEffect2} and $H_0 r_c =20$ in figure~\ref{fig:CosmoEffect3}. 
% In support of this statement, we show the response of the nDGP boost factor to the change of the 5 cosmological parameters (plots' titles), for 5 redshift values (colour legend) and different $H_0 r_c $ values.
These figures not only show that the response is limited to $\sim 7\%$ in the most extreme case (variations of $\Omega_m$ for $H_0 r_c=0.2$ at redshift $z=0$) but also that its scale dependence is similar across the different cosmological parameters and values of $H_0 r_c$. The amplitude of the effect is smaller for larger $H_0 r_c$ values.

% \begin{figure}
% \centering
% \includegraphics[width=.98\textwidth]{Figures/Emulator/nDGPboost_Ratio.pdf}
% \caption{\label{fig:BoostRatio} Ratio of the nDGP boost factors for three values of the parameter $H_0 r_c$ with respect to the boost factor of N2. The legend describes the redshift value of each line. }
% \end{figure}

%===---###===---###===---###===---###===---###===---###===---###===---###===---###===---###===---###===---###===---###===---###===---###===---###===---###===---###===---###===---###
\subsection{Simulations and data sets}
\label{ssec:SimsAndData}
We produce two independent simulation suites, one for the training set and the other for the test set. 
% Simulation settings
Both training and test set simulations use $512^3$ particles to track the evolution of the matter density field in a $(512\mpcoh)^3$ box. The forces are calculated using $1024^3$ mesh grids. The simulations are started at $z_{\rm ini}=19$ and evolved up to $z_{\rm fin}=0$ using 50 time-steps linearly spaced in the scale factor. 
% Sampling technique
We sample the cosmological parameter space using the Latin Hypercube Sampling (LHS) technique \cite{LatinHypercubeSampling}, a space-filling sampling method widely used in computer experiments to produce near-random sets of points which represent the real variability of multi-dimensional sample spaces. The range of cosmological parameters spanned by the training and test sets is described in table~\ref{tab:EmuCosmoRange}. We use 20 cosmologies\footnote{In this work we assume $k_{\rm pivot}=0.05 \, {\rm Mpc}^{-1}$ for the primordial power spectrum.} for the training set and 10 for the test set, which are listed in table~\ref{tab:TrainingCosmos} and table~\ref{tab:TestCosmos} respectively. For the modified gravity parameter $H_0 r_c$ we opt for a logarithmic sampling with 21 points in the range $[0.2,20]$ to produce the training set, and with 10 log-random points in the same range for the test set. The redshift sampling is determined by the time-stepping of COLA simulations. %, which is linear in the scale factor with 41 points between $z_{\rm max}=3$ and $z_{\rm min}=0$ for the settings used in this case.
For each cosmology in the training set (test set), we run a vanilla GR simulation and nDGP simulations in all the gravity models of the training set (test set) using the same IC. 

\begin{table}[t]
    \begin{center} 
        \begin{tabular}{lcc} 
            \toprule
            Parameter & Min. & Max. \\
            \midrule
            $h$  & $0.61$  & $0.73$ \\
            $\Omega_{\rm b}$  & $0.04$  & $0.06$ \\
            $\Omega_{\rm m}$  & $0.28$  & $0.36$ \\
            $n_{\rm s}$  & $0.92$  & $1.0$ \\
            $A_{\rm s}$  & $1.7\times 10^{-9}$  & $2.5\times 10^{-9}$ \\
            \bottomrule						
        \end{tabular}
    \end{center}
    \caption{Range of cosmological parameters spanned by the nDGP emulator.}
    \label{tab:EmuCosmoRange} 
\end{table}

% Computational complexity
Each simulation takes roughly 28 minutes on a single node of the Sciama High-Performance Compute (HPC) cluster using 32 tasks\footnote{The specific partition used to run the simulations mounts \href{https://www.intel.com/content/www/us/en/products/sku/91766/intel-xeon-processor-e52683-v4-40m-cache-2-10-ghz/specifications.html}{Intel Xeon E5-2683v4} processors.}. This corresponds to $\approx 15$ CPU hours. In total, we ran $22\times20=440$ simulations for the training set and $11\times10=110$ simulations for the test set. This means that the total computational cost for the simulations is $\approx 8 \cdot 10^3$ CPU hours. In particular, we have used a partition counting 12 nodes with 32 cores each for a total of 384 cores (a typical value for an HPC cluster) obtaining all the power spectra in less than one wall-clock day.

\begin{table}[t]
    \footnotesize
    % \centering
    \begin{tabular}{ccccc}
    \toprule
    $\Omega_{\rm m}$ &      $\Omega_{\rm b}$ &      $h$ &     $n_s$ &            $A_s$ [$\cdot 10^{-9}$]\\
    \midrule
    0.282 &  0.0515 &  0.667 &  0.950 &  2.12 \\
    0.286 &  0.0445 &  0.655 &  0.990 &  2.44 \\
    0.294 &  0.0585 &  0.697 &  0.938 &  2.00 \\
    0.298 &  0.0415 &  0.619 &  0.978 &  2.04 \\
    0.290 &  0.0545 &  0.625 &  0.986 &  1.80 \\
    0.302 &  0.0485 &  0.709 &  0.922 &  1.72 \\
    0.306 &  0.0435 &  0.649 &  0.942 &  2.36 \\
    0.314 &  0.0595 &  0.721 &  0.958 &  2.28 \\
    0.318 &  0.0575 &  0.661 &  0.962 &  2.32 \\
    0.310 &  0.0405 &  0.673 &  0.966 &  1.88 \\
    \bottomrule
    \end{tabular}
    \hfill
    \begin{tabular}{ccccc}
    \toprule
    $\Omega_{\rm m}$ &      $\Omega_{\rm b}$ &      $h$ &     $n_s$ &            $A_s$ [$\cdot 10^{-9}$]\\
    \midrule
    0.322 &  0.0465 &  0.727 &  0.982 &  1.96 \\
    0.326 &  0.0455 &  0.637 &  0.994 &  2.08 \\
    0.334 &  0.0535 &  0.631 &  0.974 &  2.16 \\
    0.338 &  0.0555 &  0.703 &  0.954 &  1.92 \\
    0.330 &  0.0525 &  0.613 &  0.934 &  2.24 \\
    0.342 &  0.0495 &  0.685 &  0.998 &  2.20 \\
    0.346 &  0.0565 &  0.691 &  0.970 &  2.48 \\
    0.354 &  0.0505 &  0.643 &  0.930 &  1.76 \\
    0.358 &  0.0425 &  0.679 &  0.926 &  2.40 \\
    0.350 &  0.0475 &  0.715 &  0.946 &  1.84 \\
    \bottomrule
    \end{tabular}
    \caption{List of cosmologies used to create the training set for the nDGP emulator.}
    \label{tab:TrainingCosmos}
\end{table}

\begin{table}[t]
    % \centering
    \footnotesize
    \begin{tabular}{ccccc}
    \toprule
    $\Omega_{\rm m}$ &      $\Omega_{\rm b}$ &      $h$ &     $n_s$ &            $A_s$ [$\cdot 10^{-9}$]\\
    \midrule
    0.284 &  0.0410 &  0.700 &  0.956 &  1.82 \\
    0.292 &  0.0490 &  0.628 &  0.940 &  1.74 \\
    0.308 &  0.0510 &  0.652 &  0.964 &  2.22 \\
    0.316 &  0.0430 &  0.688 &  0.996 &  2.30 \\
    0.324 &  0.0470 &  0.712 &  0.932 &  2.06 \\
    \bottomrule
    \end{tabular}
    \hfill
    \begin{tabular}{ccccc}
    \toprule
    $\Omega_{\rm m}$ &      $\Omega_{\rm b}$ &      $h$ &     $n_s$ &            $A_s$ [$\cdot 10^{-9}$]\\
    \midrule
    0.332 &  0.0530 &  0.724 &  0.980 &  2.38 \\
    0.348 &  0.0450 &  0.616 &  0.972 &  1.98 \\
    0.340 &  0.0550 &  0.664 &  0.948 &  1.90 \\
    0.356 &  0.0570 &  0.640 &  0.988 &  2.46 \\
    0.300 &  0.0590 &  0.676 &  0.924 &  2.14 \\
    \bottomrule
    \end{tabular}
    \caption{List of cosmologies used to create the test set for the nDGP emulator.}
    \label{tab:TestCosmos}
\end{table}

% \ToDo{Describe the technique used to compute the power spectra. Comment on the data cuts performed (i.e., k, redshift).}

The power spectra are computed by interpolating the DM distribution on a $1024^3$ mesh grid with the cloud-in-cell mass-assignment scheme. The power spectra are corrected for the window function used in the interpolation and for shot noise. We use bins of width $\Delta k= k_f$ between $k_{\rm min}=\frac{1}{2} k_f$ and $k_{\rm Nyquist}= 5 \hompc$, where $k_f$ is the fundamental frequency of the box. %As COLA simulations with these specifications have $\sim1\%$ reliable determination of the cosmological response function only up to $k \sim 1\hompc$ we cut the power spectra at $k_{\rm max}= 1 \hompc$ 
Our target accuracy is $1 \%$ up to $k= 1 \hompc$, scales up to which we have already proven COLA simulations with these specifications to have $\sim1\%$ reliable determination of the cosmological response function~\cite{Brando:2022gvg}. However for practical applications (e.g. weak lensing predictions) we extend the emulation range to $k= 5 \hompc$, where the analysis discussed in the previous section suggests an accuracy of our simulations of $\sim3\%$ for the nDGP boost factor. Beyond $k= 5 \hompc$, baryonic effects are dominant over MG effects and it is still unclear whether we can extract additional information from these small scales \cite{DES:2021wwk}.  
As we have seen that mass resolution is a limiting factor at high redshift, we make the conservative choice of restricting our data sets to the power spectra at $z_{\rm max} = 2$. We take the ratio of the power spectra in nDGP theories with the respective power spectrum in GR for each redshift and cosmology to obtain the nDGP boost factors. As the nDGP and GR simulations are run from the same IC, the sample variance is largely cancelled out when computing the nDGP boost factors. Since our simulations are started at redshift $z_{\rm ini}=19$ we multiply the nDGP boost factors with the linear theory correction in eq.~\eqref{GrowthLinCorr}.

\begin{figure}[t]
\centering 
\includegraphics[width=.48\textwidth]{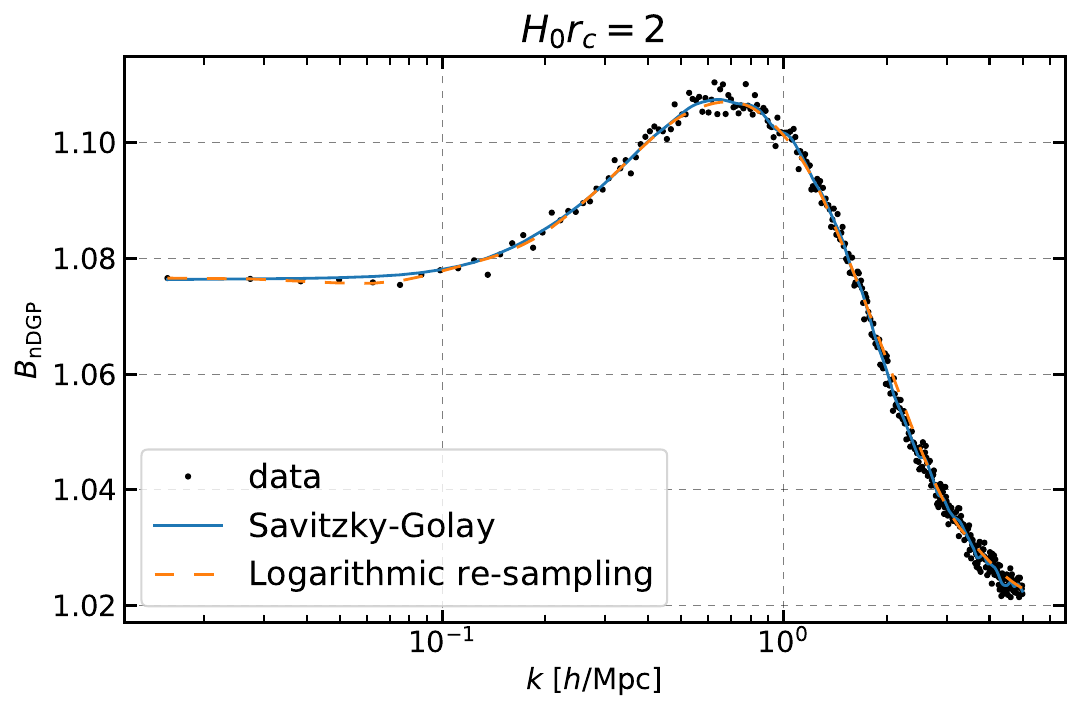}
\includegraphics[width=.48\textwidth]{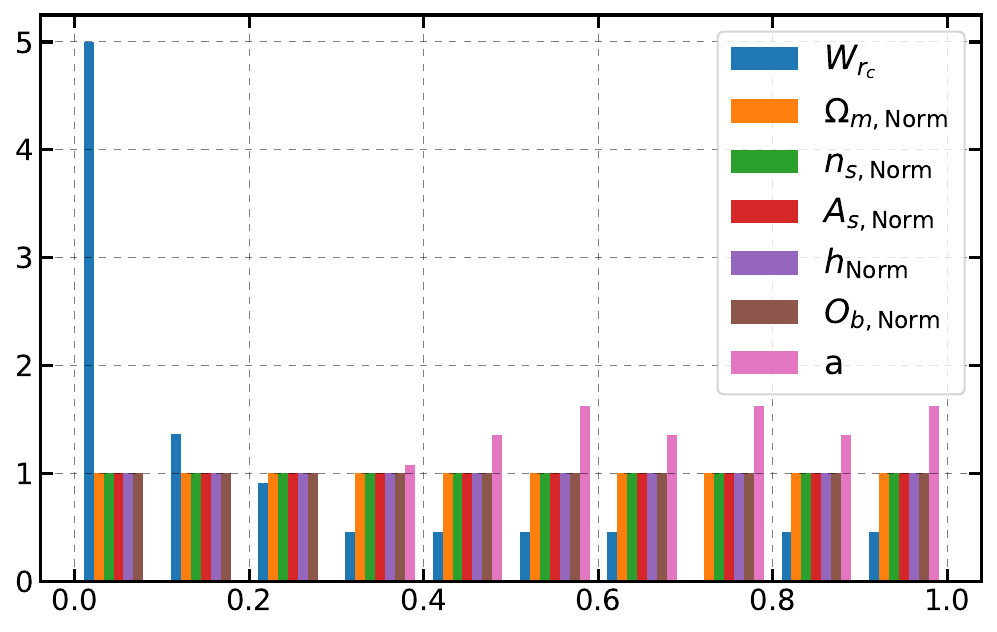}
\caption{\label{fig:SmoothAndNorm} \textit{On the left.} Comparison of the two smoothing techniques for the boost factor of nDGP with $H_0 r_c=2$. \textit{On the right.} Distribution of the normalised parameters across the training set in 10 bins between 0 and 1: all but $W_{r_c}$ and $a$ are uniformly distributed, simplifying the emulation problem.}
\end{figure}

% \subsubsection{Smoothing}\label{Sec:Smoothing}
Due to the numerical methods used to estimate the power spectrum of the {\it N}-body particles, the nDGP boost factors that we obtain are sampled with a linear binning in $k$-space and are affected by finite resolution noise at small scales. This noise is non-physical and may worsen the interpolation accuracy, therefore we decide to smooth the boost factors. We compare two smoothing techniques: one based on a logarithmic re-sampling of the data, the other based on the Savitzky-Golay filter \cite{SavitzkyGolay}. In the left plot of Fig~\ref{fig:SmoothAndNorm} is shown a comparison of the two smoothing techniques for the central case of $H_0 r_c = 2$ and $z=0$. The two techniques give consistent results so we decided to perform the rest of the analysis with the Savitzky-Golay filter. 

%===---###===---###===---###===---###===---###===---###===---###===---###===---###===---###===---###===---###===---###===---###===---###===---###===---###===---###===---###===---###
% \subsubsection{Parameters normalisation}\label{Sec:ParmsNorm}
Multi-layer perceptrons (MLP), simple neural networks that we want to employ for the parameter space interpolation, work better when they deal with normalised input values \cite{NNbook}, hence we re-scale our parameters accordingly. We map the cosmological parameters in such a way that the minimum and maximum values they assume in parameter space are 0 and 1. The redshift is naturally mapped in the range $[\frac{1}{3},1]$ by computing the scale factor. We finally convert the MG parameter $H_0 r_c$ into $W_{r_c} \equiv \frac{0.2}{H_0 r_c}$ which spans $[0.01,1]$ for our $H_0 r_c$ range (or more precisely it spans $[0,1]$ as we include the GR trivial results). In the right plot of Fig~\ref{fig:SmoothAndNorm}, the distribution of the normalised parameters is summarised in 10 linear bins between 0 and 1.

%===---###===---###===---###===---###===---###===---###===---###===---###===---###===---###===---###===---###===---###===---###===---###===---###===---###===---###===---###===---###
% \subsubsection{Logarithmic boost factor remapping}\label{Sec:LogBoostMap}
We convert the boost factor with the formula
\begin{equation}\label{B_norm}
     B_{\rm Norm}(k) = -\frac{\log_{10}(B(k) - 0.999)}{3} \, ,
    % -np.log10(df_Bk_savgol.values-(1-1e-3))/3
\end{equation}
to give more importance to small deviations from GR than to larger deviations\footnote{Notice that the boost for $W_{r_c}=0$ (limit in which we recover GR) is trivial, i.e. $B(k)=1$. In this case, we get $B_{\rm Norm}=1$ which explains the choice of the factor $1/3$ in eq~\eqref{B_norm}. The subtraction of the value $0.999$ inside the logarithm is instead performed to place the GR result at a reasonable distance from the nDGP weakest deviation result in the $B_{\rm Norm}$-space.}.
%===---###===---###===---###===---###===---###===---###===---###===---###===---###===---###===---###===---###===---###===---###===---###===---###===---###===---###===---###===---###
% \subsubsection{Principal Component Analysis}\label{Sec:PCA}
The simplicity and self-similarity of the nDGP boost factor curve suggest that the 407 $k$-modes that we are using to represent the boost factor are probably unnecessary. After subtracting the mean value of the boost factor from each element of the dataset, we perform a Principal Component Analysis (PCA) \cite{Jolliffe2002} to study if the variance of the training set is significantly larger in some $k$-modes-space directions than in others. We find that the component with the greatest variance accounts for more than $99 \%$ of the variance, while by using the first three components it is possible to describe $\sim 99.98 \%$ of the variance. This means that we can reduce the dimensionality of the problem from 407 output values (one for each $k$ bin) to 3 output values losing only $\sim 0.02 \%$ of the information. Furthermore, this remapping from the $k$-space manifold to the 3-dimensional manifold of the principal eigenvalues makes the emulation problem insensitive to the choice of $k$-binning used to compute the power spectra.
%===---###===---###===---###===---###===---###===---###===---###===---###===---###===---###===---###===---###===---###===---###===---###===---###===---###===---###===---###===---###
\subsection{Emulator}
\label{Sec:Emulator}
\begin{figure}
\centering 
\includegraphics[width=.8\textwidth]{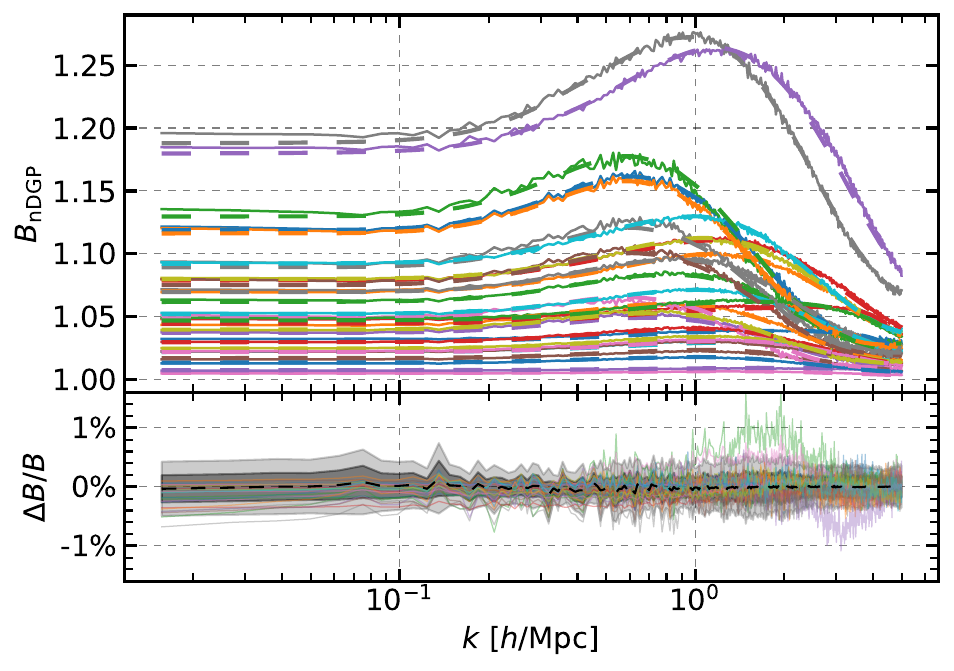}
\caption{\label{fig:EmulatorPerformance} Comparison between boost factors of the test set (solid lines) and emulated boost factors (dashed lines) for 30 examples randomly chosen in the test set. The residuals between emulated and true boost factors (coloured solid lines) are shown in the bottom panel together with the mean (dashed black line) and variance ($1 \sigma$ and $2 \sigma$ shaded regions) of the emulation error on the test-set.}
\end{figure}

We train an MLP with one hidden layer of 100 nodes activated by hyperbolic tangent functions with the limited-memory Broyden–Fletcher–Goldfarb–Shanno \cite{Fletcher1988PracticalMO} optimiser\footnote{This iterative optimisation technique exploits a numerical estimate of the Hessian based only on gradient evaluations to determine the descent direction, which makes the technique computationally more efficient than Newton's method.} on the training set. This converges in less than 5000 iterations, producing the nDGP emulator that we benchmark on the test set. In the top panel of figure~\ref{fig:EmulatorPerformance} we show some boost factors examples taken from the test set (solid lines) and compared with the emulator's predictions (dashed lines). It is worth noticing that the emulated boost factors are free of high-frequency noise thanks to the smoothing discussed in subsection~\ref{ssec:SimsAndData}. In the bottom panel of figure~\ref{fig:EmulatorPerformance} we show the mean (black dashed line) and variance ($1 \sigma$ and $2 \sigma$ shaded regions) of the emulation error on the test set together with the residuals for the 30 examples shown in the top panel. The $2 \sigma$ contours are well within the $1\%$ threshold at all scales. 

\begin{figure}
\centering 
\includegraphics[width=.8\textwidth]{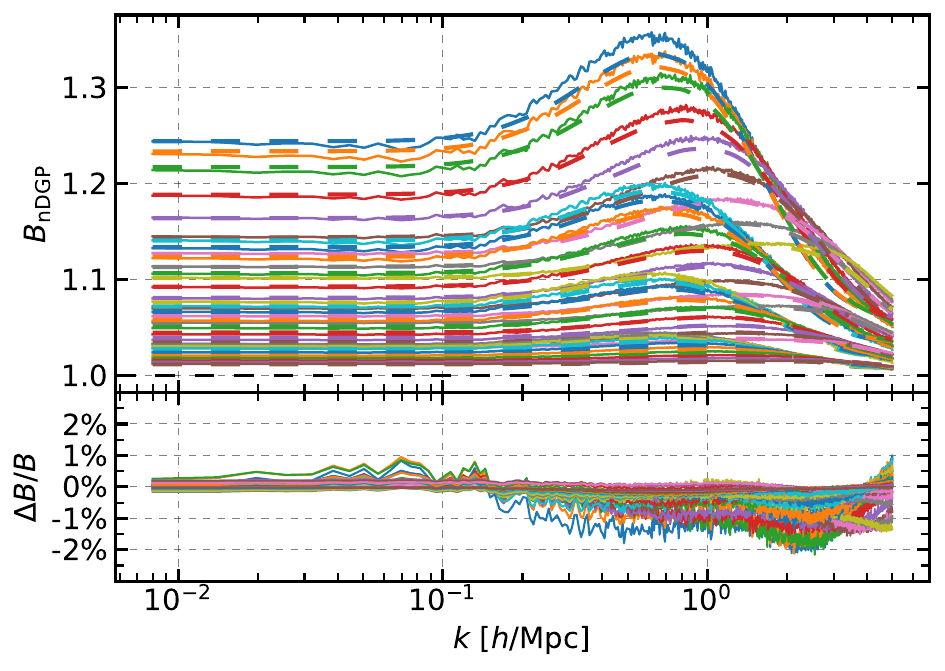}
\caption{\label{fig:EmulatorVsArepo} Comparison between emulator predictions (dashed lines) and Arepo results (solid lines) of the nDGP boost factor at 9 redshifts $z \in [0,2]$ for 4 values of the parameter $H_0 r_c$: 0.5 (in red), 1 (in orange), 2 (in blue) and 5 (in green). The bottom panel shows the residuals.}
\end{figure}

As an additional test of the emulation accuracy, we compare the emulator's predictions with the boost computed from Arepo simulations in nDGP gravity using 4 values of $H_0 r_c$: 0.5, 1, 2, 5. Thanks to the flexibility of the emulator, unlike in the comparison between PM and Arepo simulations of figure~\ref{fig:BnDGP_COLAvsArepo}, we are able in this case to compare the boosts at each of the 9 output redshifts of the Arepo simulation suite within the interpolation range. The results are shown in figure~\ref{fig:EmulatorVsArepo}. The emulator predicts the nDGP boost factor of Arepo with $\sim 2\%$ accuracy in all cases and at all the scales considered. Additionally, we have checked that the emulator performance reaches $\sim 1\%$ accuracy if restricting the redshift range to $z \in [0.5, 2]$. This is made possible by the excellent agreement between the nDGP boost factors of COLA and Arepo simulations but also by the linear-theory correction of eq~\eqref{GrowthLinCorr} applied to the nDGP boost factors of COLA simulations started at redshift $z_{\rm ini}=19$.

It is important to notice that even the results of these Arepo simulations might be not perfectly converged, but it is beyond the scope of this paper to assess the accuracy of these full \nbody{} simulations. Furthermore, we have Arepo results for specific values of the modified gravity parameter $H_0 r_c$, so the accuracy of the emulator at the edges of the parameter space cannot be fully assessed. With these caveats, and considering the level of convergence of the COLA predictions for the nDGP boost factor discussed in subsection~\ref{Sec:AccSens}, we estimate the accuracy of the emulator to be of $\sim 2\%$ up to $k=1\hompc$ and of $\sim 3\%$ up to $k=5\hompc$. Once a thorough assessment of the accuracy of the Arepo boost factors used here as ground truth should become available, this should be factored into the error budget for the emulator.

The biggest computational cost behind the nDGP emulator comes from running the COLA simulations necessary to produce the power spectra in GR and nDGP theories ($\sim 10^4$ CPU hours). The data processing applied to the power spectra in order to produce the training set has a minor impact on the computational cost ($\sim10$ CPU minutes). Also, the training of the emulator is relatively fast ($\sim1$ CPU hour). Once the model is trained, predictions are extremely fast. On an average single-core CPU, the model can produce $\sim 5000$ predictions in a second. The prediction speed of emulators is key to enabling fast cosmological inference with Monte Carlo techniques which require $10^5-10^6$ evaluations \cite{Donald-McCann:2021nxc,SpurioMancini:2021ppk, Donald-McCann:2022pac}. 

As we have shown here, DM power spectrum emulators can be extended to MG models, and in particular to nDGP gravity, with modest computational cost thanks to COLA simulations, enabling accurate and extremely fast predictions of the matter power spectrum up to $k=5 \mpcoh$.

\section{Conclusions}
\label{sec:Conclusions}
This article focused on the application of the COLA simulations for predictions of the matter power spectrum in GR and Modified Gravity (MG) theories. 
We addressed the issue of errors arising from finite resolution in numerical implementations of simulations, including time-resolution, force-resolution, and mass-resolution. Based on our analysis we designed and exploited a COLA simulation suite to extend cosmological emulators to nDGP gravity. Together with this paper, we release the nDGP emulator that will be publicly available at the address: \url{https://github.com/BartolomeoF/nDGPemu}.

To study the convergence of COLA simulations in $\Lambda$CDM, we argued that COLA converges to PM-only simulations increasing the time resolution, and referred to results obtained in this limit as PM results. 
By comparing high force-resolution PM simulations with Bacco emulator (trained with high mass-resolution simulations) and low mass-resolution \nbody{} simulations (Arepo) for predictions of the matter power spectrum at redshift $z=1$, we highlighted the importance of mass resolution, responsible for $\sim 4\%$ differences at $k=1\hompc$, while showing that PM simulations reproduce the reference results in both the low mass-resolution and high-mass-resolution cases with $\sim 1\%$ accuracy up to $k\sim 2 \hompc$. After that, we produced a suite of simulations in $\Lambda$CDM with the same cosmology but with different numbers of particles, force grids and time-steps, and using high-resolution PM simulations as a reference, we tested the relative convergence of COLA simulations. We concluded that the power spectra converge towards the reference only when mass-resolution, force-resolution and time-resolution are increased accordingly.

To showcase the potentiality of COLA for extending cosmological emulators to MG theories, we developed an nDGP boost-factor emulator of the matter power spectrum in section~\ref{sec:Boost}. 
By using Arepo predictions as a reference, we investigated the accuracy of COLA simulations for the nDGP boost factor in relation to the resolution settings and identified minimal requirements for our target accuracy. To include MG effects at $z>z_{\rm ini}$ in the back-scaling approach, we proposed and validated the linear theory correction of eq~\eqref{GrowthLinCorr}.
We also studied the response of the nDGP boost factors to the change of cosmological parameters finding weak cosmological dependence of the nDGP boost factor and confirming that the cosmological dependence becomes weaker for smaller deviations from $\Lambda$CDM (larger values of $H_0 r_c$). In subsection~\ref{ssec:SimsAndData}, we detailed our simulation strategy to compute the nDGP boost factors and examined the computational cost consisting of a total of $\approx 8 \cdot 10^{3}$ CPU hours. 
After smoothing the signals and remapping the parameters in a unitary interval, we performed a PCA and selected only 3 components while preserving $99.98 \%$ of the information. We then used the resulting data set to train a neural network and bench-marked its accuracy on an independent test set ($<1\%$ at all scales considered). As an additional test, we compared the emulator predictions with the Arepo results and found better than $2\%$ agreement at all scales considered. We concluded that COLA simulations can be used to accurately extend matter power spectrum emulators to MG theories up to $k=1\hompc$ ($k=5\hompc$) with $2\%$ accuracy ($3\%$ accuracy).

The approach developed in this work can be extended to other gravity models. In particular it would be interesting to train an emulator for the matter power spectrum in Horndeski gravity~\cite{Horndeski:1974wa,Deffayet:2011gz} which is made possible thanks to the recent extension of the COLA method \cite{Wright:2022krq,Brando:2022gvg,Brando:2023fzu} to the sub-class of Horndeski theories incorporating the Vainshtein screening mechanism.
Furthermore, we plan to extend the convergence of COLA simulations for predictions of halo statistics for which we have already collected promising results. Finally, we aim to develop a suite of COLA-based emulators for several LSS probes, such as the ones studied in~\cite{Fiorini:2021dzs,Fiorini:2022srj}.

\acknowledgments
We thank Albert Izard, Daniela Saadeh and Jamie Donald-McCann for useful insight and stimulating discussions on COLA simulations and cosmological emulators. We thank Hans Winther for creating the FML library and continuously improving it  with new features. We thank César Hernández-Aguayo for providing access to the MG-Arepo simulations. 
For the purpose of open access, the authors have applied a Creative Commons Attribution (CC BY) licence to any Author Accepted Manuscript version arising from this work. 
Numerical computations were done on the Sciama High Performance Compute (HPC) cluster which is supported by the ICG, SEPNet and the University of Portsmouth. BF is supported by a Royal Society Enhancement Award (grant no. RF$\backslash$ERE$\backslash$210304). KK is supported by the STFC grant ST/W001225/1. TB is supported by ERC Starting Grant \textit{SHADE} (grant no.~StG 949572) and a Royal Society University Research Fellowship (grant no.~URF$\backslash$R1$\backslash$180009).

\paragraph{Data availability} Supporting research data are available on reasonable request from the corresponding author.

% Bibliography

% [A] Recommended: using JHEP.bst file
\bibliographystyle{JHEP}
\bibliography{biblio.bib}

\providecommand{\href}[2]{#2}\begingroup\raggedright\begin{thebibliography}{10}

\bibitem{DESI:2023dwi}
{\scshape DESI} collaboration, \emph{{Validation of the Scientific Program for
  the Dark Energy Spectroscopic Instrument}},
  \href{https://arxiv.org/abs/2306.06307}{{\ttfamily 2306.06307}}.

\bibitem{Chisari:2019tus}
N.E.~Chisari et~al., \emph{{Modelling baryonic feedback for survey cosmology}},
  \href{https://doi.org/10.21105/astro.1905.06082}{\emph{Open J. Astrophys.}
  {\bfseries 2} (2019) 4} [\href{https://arxiv.org/abs/1905.06082}{{\ttfamily
  1905.06082}}].

\bibitem{DES:2021wwk}
{\scshape DES} collaboration, \emph{{Dark Energy Survey Year 3 results:
  Cosmological constraints from galaxy clustering and weak lensing}},
  \href{https://doi.org/10.1103/PhysRevD.105.023520}{\emph{Phys. Rev. D}
  {\bfseries 105} (2022) 023520}
  [\href{https://arxiv.org/abs/2105.13549}{{\ttfamily 2105.13549}}].

\bibitem{Lange:2023khv}
J.U.~Lange, A.P.~Hearin, A.~Leauthaud, F.C.~van~den Bosch, E.~Xhakaj, H.~Guo
  et~al., \emph{{Constraints on $S_8$ from a full-scale and full-shape analysis
  of redshift-space clustering and galaxy-galaxy lensing in BOSS}},
  \href{https://arxiv.org/abs/2301.08692}{{\ttfamily 2301.08692}}.

\bibitem{Tassev:2013pn}
S.~Tassev, M.~Zaldarriaga and D.~Eisenstein, \emph{{Solving Large Scale
  Structure in Ten Easy Steps with COLA}},
  \href{https://doi.org/10.1088/1475-7516/2013/06/036}{\emph{JCAP} {\bfseries
  06} (2013) 036} [\href{https://arxiv.org/abs/1301.0322}{{\ttfamily
  1301.0322}}].

\bibitem{Izard:2015dja}
A.~Izard, M.~Crocce and P.~Fosalba, \emph{{ICE-COLA: Towards fast and accurate
  synthetic galaxy catalogues optimizing a quasi $N$-body method}},
  \href{https://doi.org/10.1093/mnras/stw797}{\emph{Mon. Not. Roy. Astron.
  Soc.} {\bfseries 459} (2016) 2327}
  [\href{https://arxiv.org/abs/1509.04685}{{\ttfamily 1509.04685}}].

\bibitem{List:2023jxz}
F.~List and O.~Hahn, \emph{{Perturbation-theory informed integrators for
  cosmological simulations}},
  \href{https://arxiv.org/abs/2301.09655}{{\ttfamily 2301.09655}}.

\bibitem{Koda:2015mca}
J.~Koda, C.~Blake, F.~Beutler, E.~Kazin and F.~Marin, \emph{{Fast and accurate
  mock catalogue generation for low-mass galaxies}},
  \href{https://doi.org/10.1093/mnras/stw763}{\emph{Mon. Not. Roy. Astron.
  Soc.} {\bfseries 459} (2016) 2118}
  [\href{https://arxiv.org/abs/1507.05329}{{\ttfamily 1507.05329}}].

\bibitem{Drinkwater:2009sd}
M.J.~Drinkwater et~al., \emph{{The WiggleZ Dark Energy Survey: Survey Design
  and First Data Release}},
  \href{https://doi.org/10.1111/j.1365-2966.2009.15754.x}{\emph{Mon. Not. Roy.
  Astron. Soc.} {\bfseries 401} (2010) 1429}
  [\href{https://arxiv.org/abs/0911.4246}{{\ttfamily 0911.4246}}].

\bibitem{gadget2}
V.~Springel, \emph{{The Cosmological simulation code GADGET-2}},
  \href{https://doi.org/10.1111/j.1365-2966.2005.09655.x}{\emph{Mon. Not. Roy.
  Astron. Soc.} {\bfseries 364} (2005) 1105}
  [\href{https://arxiv.org/abs/astro-ph/0505010}{{\ttfamily
  astro-ph/0505010}}].

\bibitem{Howlett:2015hfa}
C.~Howlett, M.~Manera and W.J.~Percival, \emph{{L-PICOLA: A parallel code for
  fast dark matter simulation}},
  \href{https://doi.org/10.1016/j.ascom.2015.07.003}{\emph{Astron. Comput.}
  {\bfseries 12} (2015) 109}
  [\href{https://arxiv.org/abs/1506.03737}{{\ttfamily 1506.03737}}].

\bibitem{Izard:2017kma}
A.~Izard, P.~Fosalba and M.~Crocce, \emph{{ICE-COLA: fast simulations for weak
  lensing observables}},
  \href{https://doi.org/10.1093/mnras/stx2544}{\emph{Mon. Not. Roy. Astron.
  Soc.} {\bfseries 473} (2018) 3051}
  [\href{https://arxiv.org/abs/1707.06312}{{\ttfamily 1707.06312}}].

\bibitem{Valogiannis:2016ane}
G.~Valogiannis and R.~Bean, \emph{{Efficient simulations of large scale
  structure in modified gravity cosmologies with comoving Lagrangian
  acceleration}}, \href{https://doi.org/10.1103/PhysRevD.95.103515}{\emph{Phys.
  Rev. D} {\bfseries 95} (2017) 103515}
  [\href{https://arxiv.org/abs/1612.06469}{{\ttfamily 1612.06469}}].

\bibitem{Winther:2017jof}
H.A.~Winther, K.~Koyama, M.~Manera, B.S.~Wright and G.-B.~Zhao, \emph{{COLA
  with scale-dependent growth: applications to screened modified gravity
  models}}, \href{https://doi.org/10.1088/1475-7516/2017/08/006}{\emph{JCAP}
  {\bfseries 08} (2017) 006}
  [\href{https://arxiv.org/abs/1703.00879}{{\ttfamily 1703.00879}}].

\bibitem{Wright:2022krq}
{\scshape LSST Dark Energy Science} collaboration, \emph{{Hi-COLA: fast,
  approximate simulations of structure formation in Horndeski gravity}},
  \href{https://doi.org/10.1088/1475-7516/2023/03/040}{\emph{JCAP} {\bfseries
  03} (2023) 040} [\href{https://arxiv.org/abs/2209.01666}{{\ttfamily
  2209.01666}}].

\bibitem{Fiorini:2021dzs}
B.~Fiorini, K.~Koyama, A.~Izard, H.A.~Winther, B.S.~Wright and B.~Li,
  \emph{{Fast generation of mock galaxy catalogues in modified gravity models
  with COLA}}, \href{https://doi.org/10.1088/1475-7516/2021/09/021}{\emph{JCAP}
  {\bfseries 09} (2021) 021}
  [\href{https://arxiv.org/abs/2106.05197}{{\ttfamily 2106.05197}}].

\bibitem{Wright:2017dkw}
B.S.~Wright, H.A.~Winther and K.~Koyama, \emph{{COLA with massive neutrinos}},
  \href{https://doi.org/10.1088/1475-7516/2017/10/054}{\emph{JCAP} {\bfseries
  10} (2017) 054} [\href{https://arxiv.org/abs/1705.08165}{{\ttfamily
  1705.08165}}].

\bibitem{Winther:2014cia}
H.A.~Winther and P.G.~Ferreira, \emph{{Fast route to nonlinear clustering
  statistics in modified gravity theories}},
  \href{https://doi.org/10.1103/PhysRevD.91.123507}{\emph{Phys. Rev. D}
  {\bfseries 91} (2015) 123507}
  [\href{https://arxiv.org/abs/1403.6492}{{\ttfamily 1403.6492}}].

\bibitem{Scoccimarro:2009eu}
R.~Scoccimarro, \emph{{Large-Scale Structure in Brane-Induced Gravity I.
  Perturbation Theory}},
  \href{https://doi.org/10.1103/PhysRevD.80.104006}{\emph{Phys. Rev. D}
  {\bfseries 80} (2009) 104006}
  [\href{https://arxiv.org/abs/0906.4545}{{\ttfamily 0906.4545}}].

\bibitem{Brando:2023fzu}
G.~Brando, K.~Koyama and H.A.~Winther, \emph{{Revisiting Vainshtein Screening
  for fast N-body simulations}},
  \href{https://arxiv.org/abs/2303.09549}{{\ttfamily 2303.09549}}.

\bibitem{agarwal2012}
S.~{Agarwal}, F.B.~{Abdalla}, H.A.~{Feldman}, O.~{Lahav} and S.A.~{Thomas},
  \emph{{PkANN - I. Non-linear matter power spectrum interpolation through
  artificial neural networks}},
  \href{https://doi.org/10.1111/j.1365-2966.2012.21326.x}{\emph{\mnras}
  {\bfseries 424} (2012) 1409}
  [\href{https://arxiv.org/abs/1203.1695}{{\ttfamily 1203.1695}}].

\bibitem{habib2007}
S.~Habib, K.~Heitmann, D.~Higdon, C.~Nakhleh and B.~Williams, \emph{{Cosmic
  Calibration: Constraints from the Matter Power Spectrum and the Cosmic
  Microwave Background}},
  \href{https://doi.org/10.1103/PhysRevD.76.083503}{\emph{Phys. Rev. D}
  {\bfseries 76} (2007) 083503}
  [\href{https://arxiv.org/abs/astro-ph/0702348}{{\ttfamily
  astro-ph/0702348}}].

\bibitem{Smith:2002dz}
{\scshape VIRGO Consortium} collaboration, \emph{{Stable clustering, the halo
  model and nonlinear cosmological power spectra}},
  \href{https://doi.org/10.1046/j.1365-8711.2003.06503.x}{\emph{Mon. Not. Roy.
  Astron. Soc.} {\bfseries 341} (2003) 1311}
  [\href{https://arxiv.org/abs/astro-ph/0207664}{{\ttfamily
  astro-ph/0207664}}].

\bibitem{Euclid:2020rfv}
{\scshape Euclid} collaboration, \emph{{Euclid preparation: IX. EuclidEmulator2
  \textendash{} power spectrum emulation with massive neutrinos and
  self-consistent dark energy perturbations}},
  \href{https://doi.org/10.1093/mnras/stab1366}{\emph{Mon. Not. Roy. Astron.
  Soc.} {\bfseries 505} (2021) 2840}
  [\href{https://arxiv.org/abs/2010.11288}{{\ttfamily 2010.11288}}].

\bibitem{Angulo:2020vky}
R.E.~Angulo, M.~Zennaro, S.~Contreras, G.~Aric\`o, M.~Pellejero-Iba\~nez and
  J.~St\"ucker, \emph{{The BACCO simulation project: exploiting the full power
  of large-scale structure for cosmology}},
  \href{https://doi.org/10.1093/mnras/stab2018}{\emph{Mon. Not. Roy. Astron.
  Soc.} {\bfseries 507} (2021) 5869}
  [\href{https://arxiv.org/abs/2004.06245}{{\ttfamily 2004.06245}}].

\bibitem{Winther:2019mus}
H.~Winther, S.~Casas, M.~Baldi, K.~Koyama, B.~Li, L.~Lombriser et~al.,
  \emph{{Emulators for the nonlinear matter power spectrum beyond
  $\Lambda$CDM}},
  \href{https://doi.org/10.1103/PhysRevD.100.123540}{\emph{Phys. Rev. D}
  {\bfseries 100} (2019) 123540}
  [\href{https://arxiv.org/abs/1903.08798}{{\ttfamily 1903.08798}}].

\bibitem{Ramachandra:2020lue}
{\scshape LSST Dark Energy Science} collaboration, \emph{{Matter Power Spectrum
  Emulator for f(R) Modified Gravity Cosmologies}},
  \href{https://doi.org/10.1103/PhysRevD.103.123525}{\emph{Phys. Rev. D}
  {\bfseries 103} (2021) 123525}
  [\href{https://arxiv.org/abs/2010.00596}{{\ttfamily 2010.00596}}].

\bibitem{Arnold:2021xtm}
C.~Arnold, B.~Li, B.~Giblin, J.~Harnois-D\'eraps and Y.-C.~Cai, \emph{{FORGE --
  the f(R) gravity cosmic emulator project I: Introduction and matter power
  spectrum emulator}},  \href{https://arxiv.org/abs/2109.04984}{{\ttfamily
  2109.04984}}.

\bibitem{Saez-Casares:2023olw}
I.n.~S\'aez-Casares, Y.~Rasera and B.~Li, \emph{{The e-MANTIS emulator: fast
  predictions of the non-linear matter power spectrum in $f(R)$CDM cosmology}},
   \href{https://arxiv.org/abs/2303.08899}{{\ttfamily 2303.08899}}.

\bibitem{Mauland:2023pjt}
R.~Mauland, H.A.~Winther and C.-Z.~Ruan, \emph{{Sesame: A power spectrum
  emulator pipeline for beyond-$\Lambda$CDM models}},
  \href{https://arxiv.org/abs/2309.13295}{{\ttfamily 2309.13295}}.

\bibitem{Hu:2007nk}
W.~Hu and I.~Sawicki, \emph{{Models of f(R) Cosmic Acceleration that Evade
  Solar-System Tests}},
  \href{https://doi.org/10.1103/PhysRevD.76.064004}{\emph{Phys. Rev. D}
  {\bfseries 76} (2007) 064004}
  [\href{https://arxiv.org/abs/0705.1158}{{\ttfamily 0705.1158}}].

\bibitem{Harnois-Deraps:2022bie}
J.~Harnois-D\'eraps, C.~Hernandez-Aguayo, C.~Cuesta-Lazaro, C.~Arnold, B.~Li,
  C.T.~Davies et~al., \emph{{mglens: Modified gravity weak lensing simulations
  for emulation-based cosmological inference}},
  \href{https://doi.org/10.1093/mnras/stad2700}{\emph{Mon. Not. Roy. Astron.
  Soc.} {\bfseries 525} (2023) 6336}
  [\href{https://arxiv.org/abs/2211.05779}{{\ttfamily 2211.05779}}].

\bibitem{Dvali:2000hr}
G.R.~Dvali, G.~Gabadadze and M.~Porrati, \emph{{4-D gravity on a brane in 5-D
  Minkowski space}},
  \href{https://doi.org/10.1016/S0370-2693(00)00669-9}{\emph{Phys. Lett. B}
  {\bfseries 485} (2000) 208}
  [\href{https://arxiv.org/abs/hep-th/0005016}{{\ttfamily hep-th/0005016}}].

\bibitem{Cataneo:2018cic}
M.~Cataneo, L.~Lombriser, C.~Heymans, A.~Mead, A.~Barreira, S.~Bose et~al.,
  \emph{{On the road to percent accuracy: non-linear reaction of the matter
  power spectrum to dark energy and modified gravity}},
  \href{https://doi.org/10.1093/mnras/stz1836}{\emph{Mon. Not. Roy. Astron.
  Soc.} {\bfseries 488} (2019) 2121}
  [\href{https://arxiv.org/abs/1812.05594}{{\ttfamily 1812.05594}}].

\bibitem{Bose:2022vwi}
B.~Bose, M.~Tsedrik, J.~Kennedy, L.~Lombriser, A.~Pourtsidou and A.~Taylor,
  \emph{{Fast and accurate predictions of the nonlinear matter power spectrum
  for general models of Dark Energy and Modified Gravity}},
  \href{https://arxiv.org/abs/2210.01094}{{\ttfamily 2210.01094}}.

\bibitem{Gupta:2023rbf}
S.~Gupta, W.A.~Hellwing and M.~Bilicki, \emph{{Improved analytical modeling of
  the nonlinear power spectrum in modified gravity cosmologies}},
  \href{https://doi.org/10.1103/PhysRevD.107.083525}{\emph{Phys. Rev. D}
  {\bfseries 107} (2023) 083525}
  [\href{https://arxiv.org/abs/2301.12016}{{\ttfamily 2301.12016}}].

\bibitem{Winther:2015wla}
H.A.~Winther et~al., \emph{{Modified Gravity N-body Code Comparison Project}},
  \href{https://doi.org/10.1093/mnras/stv2253}{\emph{Mon. Not. Roy. Astron.
  Soc.} {\bfseries 454} (2015) 4208}
  [\href{https://arxiv.org/abs/1506.06384}{{\ttfamily 1506.06384}}].

\bibitem{Klypin:2017iwu}
A.~Klypin and F.~Prada, \emph{{Dark matter statistics for large galaxy
  catalogues: power spectra and covariance matrices}},
  \href{https://doi.org/10.1093/mnras/sty1340}{\emph{Mon. Not. Roy. Astron.
  Soc.} {\bfseries 478} (2018) 4602}
  [\href{https://arxiv.org/abs/1701.05690}{{\ttfamily 1701.05690}}].

\bibitem{DeRose:2018xdj}
J.~DeRose, R.H.~Wechsler, J.L.~Tinker, M.R.~Becker, Y.-Y.~Mao, T.~McClintock
  et~al., \emph{{The Aemulus Project I: Numerical Simulations for Precision
  Cosmology}}, \href{https://doi.org/10.3847/1538-4357/ab1085}{\emph{Astrophys.
  J.} {\bfseries 875} (2019) 69}
  [\href{https://arxiv.org/abs/1804.05865}{{\ttfamily 1804.05865}}].

\bibitem{Fiorini:2022srj}
B.~Fiorini, K.~Koyama and A.~Izard, \emph{{Studying large-scale structure
  probes of modified gravity with COLA}},
  \href{https://arxiv.org/abs/2208.01345}{{\ttfamily 2208.01345}}.

\bibitem{Hernandez-Aguayo:2021kuh}
C.~Hern\'andez-Aguayo, C.-Z.~Ruan, B.~Li, C.~Arnold, C.M.~Baugh, A.~Klypin
  et~al., \emph{{Fast full N-body simulations of generic modified gravity:
  derivative coupling models}},
  \href{https://doi.org/10.1088/1475-7516/2022/01/048}{\emph{JCAP} {\bfseries
  01} (2022) 048} [\href{https://arxiv.org/abs/2110.00566}{{\ttfamily
  2110.00566}}].

\bibitem{Brando:2022gvg}
G.~Brando, B.~Fiorini, K.~Koyama and H.A.~Winther, \emph{{Enabling matter power
  spectrum emulation in beyond-\ensuremath{\Lambda}CDM cosmologies with COLA}},
  \href{https://doi.org/10.1088/1475-7516/2022/09/051}{\emph{JCAP} {\bfseries
  09} (2022) 051} [\href{https://arxiv.org/abs/2203.11120}{{\ttfamily
  2203.11120}}].

\bibitem{Arepo_fR}
C.~Arnold, M.~Leo and B.~Li, \emph{{Realistic simulations of galaxy formation
  in $f(R)$ modified gravity}},
  \href{https://doi.org/10.1038/s41550-019-0823-y}{\emph{Nature Astron.}
  {\bfseries 3} (2019) 945} [\href{https://arxiv.org/abs/1907.02977}{{\ttfamily
  1907.02977}}].

\bibitem{Arepo_nDGP}
C.~Hern\'andez-Aguayo, C.~Arnold, B.~Li and C.M.~Baugh, \emph{{Galaxy formation
  in the brane world I: overview and first results}},
  \href{https://doi.org/10.1093/mnras/stab694}{\emph{Mon. Not. Roy. Astron.
  Soc.} {\bfseries 503} (2021) 3867}
  [\href{https://arxiv.org/abs/2006.15467}{{\ttfamily 2006.15467}}].

\bibitem{Arepo}
V.~Springel, \emph{{E pur si muove: Galiliean-invariant cosmological
  hydrodynamical simulations on a moving mesh}},
  \href{https://doi.org/10.1111/j.1365-2966.2009.15715.x}{\emph{Mon. Not. Roy.
  Astron. Soc.} {\bfseries 401} (2010) 791}
  [\href{https://arxiv.org/abs/0901.4107}{{\ttfamily 0901.4107}}].

\bibitem{Mitchell:2021aex}
M.A.~Mitchell, C.~Hern\'andez-Aguayo, C.~Arnold and B.~Li, \emph{{A general
  framework to test gravity using galaxy clusters IV: cluster and halo
  properties in DGP gravity}},
  \href{https://doi.org/10.1093/mnras/stab2817}{\emph{Mon. Not. Roy. Astron.
  Soc.} {\bfseries 508} (2021) 4140}
  [\href{https://arxiv.org/abs/2106.13815}{{\ttfamily 2106.13815}}].

\bibitem{Euclid:2019clj}
{\scshape Euclid} collaboration, \emph{{Euclid preparation: VII. Forecast
  validation for Euclid cosmological probes}},
  \href{https://doi.org/10.1051/0004-6361/202038071}{\emph{Astron. Astrophys.}
  {\bfseries 642} (2020) A191}
  [\href{https://arxiv.org/abs/1910.09273}{{\ttfamily 1910.09273}}].

\bibitem{Michaux:2020yis}
M.~Michaux, O.~Hahn, C.~Rampf and R.E.~Angulo, \emph{{Accurate initial
  conditions for cosmological N-body simulations: Minimizing truncation and
  discreteness errors}},
  \href{https://doi.org/10.1093/mnras/staa3149}{\emph{Mon. Not. Roy. Astron.
  Soc.} {\bfseries 500} (2020) 663}
  [\href{https://arxiv.org/abs/2008.09588}{{\ttfamily 2008.09588}}].

\bibitem{List:2023kbb}
F.~List, O.~Hahn and C.~Rampf, \emph{{Fluid-limit Cosmological Simulations
  Starting from the Big Bang}},
  \href{https://arxiv.org/abs/2309.10865}{{\ttfamily 2309.10865}}.

\bibitem{Feng:2016yqz}
Y.~Feng, M.-Y.~Chu, U.~Seljak and P.~McDonald, \emph{{FastPM: a new scheme for
  fast simulations of dark matter and haloes}},
  \href{https://doi.org/10.1093/mnras/stw2123}{\emph{Mon. Not. Roy. Astron.
  Soc.} {\bfseries 463} (2016) 2273}
  [\href{https://arxiv.org/abs/1603.00476}{{\ttfamily 1603.00476}}].

\bibitem{Deffayet:2000uy}
C.~Deffayet, \emph{{Cosmology on a brane in Minkowski bulk}},
  \href{https://doi.org/10.1016/S0370-2693(01)00160-5}{\emph{Phys. Lett. B}
  {\bfseries 502} (2001) 199}
  [\href{https://arxiv.org/abs/hep-th/0010186}{{\ttfamily hep-th/0010186}}].

\bibitem{Luty:2003vm}
M.A.~Luty, M.~Porrati and R.~Rattazzi, \emph{{Strong interactions and stability
  in the DGP model}},
  \href{https://doi.org/10.1088/1126-6708/2003/09/029}{\emph{JHEP} {\bfseries
  09} (2003) 029} [\href{https://arxiv.org/abs/hep-th/0303116}{{\ttfamily
  hep-th/0303116}}].

\bibitem{Koyama:2007ih}
K.~Koyama and F.P.~Silva, \emph{{Non-linear interactions in a cosmological
  background in the DGP braneworld}},
  \href{https://doi.org/10.1103/PhysRevD.75.084040}{\emph{Phys. Rev. D}
  {\bfseries 75} (2007) 084040}
  [\href{https://arxiv.org/abs/hep-th/0702169}{{\ttfamily hep-th/0702169}}].

\bibitem{Koyama:2007za}
K.~Koyama, \emph{{Ghosts in the self-accelerating universe}},
  \href{https://doi.org/10.1088/0264-9381/24/24/R01}{\emph{Class. Quant. Grav.}
  {\bfseries 24} (2007) R231}
  [\href{https://arxiv.org/abs/0709.2399}{{\ttfamily 0709.2399}}].

\bibitem{Schmidt:2009sg}
F.~Schmidt, \emph{{Self-Consistent Cosmological Simulations of DGP Braneworld
  Gravity}}, \href{https://doi.org/10.1103/PhysRevD.80.043001}{\emph{Phys. Rev.
  D} {\bfseries 80} (2009) 043001}
  [\href{https://arxiv.org/abs/0905.0858}{{\ttfamily 0905.0858}}].

\bibitem{Lombriser:2009xg}
L.~Lombriser, W.~Hu, W.~Fang and U.~Seljak, \emph{{Cosmological Constraints on
  DGP Braneworld Gravity with Brane Tension}},
  \href{https://doi.org/10.1103/PhysRevD.80.063536}{\emph{Phys. Rev. D}
  {\bfseries 80} (2009) 063536}
  [\href{https://arxiv.org/abs/0905.1112}{{\ttfamily 0905.1112}}].

\bibitem{Deffayet:2001uk}
C.~Deffayet, G.R.~Dvali, G.~Gabadadze and A.I.~Vainshtein,
  \emph{{Nonperturbative continuity in graviton mass versus perturbative
  discontinuity}},
  \href{https://doi.org/10.1103/PhysRevD.65.044026}{\emph{Phys. Rev. D}
  {\bfseries 65} (2002) 044026}
  [\href{https://arxiv.org/abs/hep-th/0106001}{{\ttfamily hep-th/0106001}}].

\bibitem{Li:2011vk}
B.~Li, G.-B.~Zhao, R.~Teyssier and K.~Koyama, \emph{{ECOSMOG: An Efficient Code
  for Simulating Modified Gravity}},
  \href{https://doi.org/10.1088/1475-7516/2012/01/051}{\emph{JCAP} {\bfseries
  01} (2012) 051} [\href{https://arxiv.org/abs/1110.1379}{{\ttfamily
  1110.1379}}].

\bibitem{Hernandez-Aguayo:2020kgq}
C.~Hern\'andez-Aguayo, C.~Arnold, B.~Li and C.M.~Baugh, \emph{{Galaxy formation
  in the brane world I: overview and first results}},
  \href{https://doi.org/10.1093/mnras/stab694}{\emph{Mon. Not. Roy. Astron.
  Soc.} {\bfseries 503} (2021) 3867}
  [\href{https://arxiv.org/abs/2006.15467}{{\ttfamily 2006.15467}}].

\bibitem{winther15}
H.A.~{Winther} and P.G.~{Ferreira}, \emph{{Fast route to nonlinear clustering
  statistics in modified gravity theories}},
  \href{https://doi.org/10.1103/PhysRevD.91.123507}{\emph{\prd} {\bfseries 91}
  (2015) 123507} [\href{https://arxiv.org/abs/1403.6492}{{\ttfamily
  1403.6492}}].

\bibitem{Fiorini:2023uzw}
B.~Fiorini, \emph{{Testing gravity on cosmological scales : theoretical
  predictions with the COLA method}}, Ph.D. thesis, University of Portsmouth,
  2023.
\newblock \href{https://arxiv.org/abs/2303.07121}{{\ttfamily 2303.07121}}.

\bibitem{Ruan:2021wup}
C.-Z.~Ruan, C.~Hern\'andez-Aguayo, B.~Li, C.~Arnold, C.M.~Baugh, A.~Klypin
  et~al., \emph{{Fast full N-body simulations of generic modified gravity:
  conformal coupling models}},
  \href{https://doi.org/10.1088/1475-7516/2022/05/018}{\emph{JCAP} {\bfseries
  05} (2022) 018} [\href{https://arxiv.org/abs/2110.00328}{{\ttfamily
  2110.00328}}].

\bibitem{LatinHypercubeSampling}
M.D.~McKay, R.J.~Beckman and W.J.~Conover, \emph{Comparison of three methods
  for selecting values of input variables in the analysis of output from a
  computer code},
  \href{https://doi.org/10.1080/00401706.1979.10489755}{\emph{Technometrics}
  {\bfseries 21} (1979) 239}
  [\href{https://arxiv.org/abs/https://doi.org/10.1080/00401706.1979.10489755}{{\ttfamily
  https://doi.org/10.1080/00401706.1979.10489755}}].

\bibitem{SavitzkyGolay}
A.~Savitzky and M.J.E.~Golay, \emph{Smoothing and differentiation of data by
  simplified least squares procedures.},
  \href{https://doi.org/10.1021/ac60214a047}{\emph{Analytical Chemistry}
  {\bfseries 36} (1964) 1627}
  [\href{https://arxiv.org/abs/https://doi.org/10.1021/ac60214a047}{{\ttfamily
  https://doi.org/10.1021/ac60214a047}}].

\bibitem{NNbook}
C.M.~Bishop, \emph{Neural Networks for Pattern Recognition}, Oxford University
  Press, Inc., USA (1995).

\bibitem{Jolliffe2002}
I.T.~Jolliffe, \emph{Principal Component Analysis}, Springer Series in
  Statistics, Springer-Verlag (2002),
  \href{https://doi.org/10.1007/b98835}{10.1007/b98835}.

\bibitem{Fletcher1988PracticalMO}
R.~Fletcher, \emph{Practical Methods of Optimization}, Wiley (2000).

\bibitem{Donald-McCann:2021nxc}
J.~Donald-McCann, F.~Beutler, K.~Koyama and M.~Karamanis, \emph{{matryoshka:
  halo model emulator for the galaxy power spectrum}},
  \href{https://doi.org/10.1093/mnras/stac239}{\emph{Mon. Not. Roy. Astron.
  Soc.} {\bfseries 511} (2022) 3768}
  [\href{https://arxiv.org/abs/2109.15236}{{\ttfamily 2109.15236}}].

\bibitem{SpurioMancini:2021ppk}
A.~Spurio~Mancini, D.~Piras, J.~Alsing, B.~Joachimi and M.P.~Hobson,
  \emph{{CosmoPower: emulating cosmological power spectra for accelerated
  Bayesian inference from next-generation surveys}},
  \href{https://doi.org/10.1093/mnras/stac064}{\emph{Mon. Not. Roy. Astron.
  Soc.} {\bfseries 511} (2022) 1771}
  [\href{https://arxiv.org/abs/2106.03846}{{\ttfamily 2106.03846}}].

\bibitem{Donald-McCann:2022pac}
J.~Donald-McCann, K.~Koyama and F.~Beutler, \emph{{$\texttt{matryoshka}$ II:
  Accelerating Effective Field Theory Analyses of the Galaxy Power Spectrum}},
  \href{https://arxiv.org/abs/2202.07557}{{\ttfamily 2202.07557}}.

\bibitem{Horndeski:1974wa}
G.W.~Horndeski, \emph{{Second-order scalar-tensor field equations in a
  four-dimensional space}},
  \href{https://doi.org/10.1007/BF01807638}{\emph{Int. J. Theor. Phys.}
  {\bfseries 10} (1974) 363}.

\bibitem{Deffayet:2011gz}
C.~Deffayet, X.~Gao, D.A.~Steer and G.~Zahariade, \emph{{From k-essence to
  generalised Galileons}},
  \href{https://doi.org/10.1103/PhysRevD.84.064039}{\emph{Phys. Rev. D}
  {\bfseries 84} (2011) 064039}
  [\href{https://arxiv.org/abs/1103.3260}{{\ttfamily 1103.3260}}].

\end{thebibliography}\endgroup

\end{document}